# Modeling dependent degradation data considering inherent causal relationships for reliability analysis and remaining useful life prediction

Shi-Shun Chen [a], Xiao-Yang Li [a, *]

[a] School of Reliability and Systems Engineering, Beihang University, Beijing, China

Email:
css1107@buaa.edu.cn (Shi-Shun Chen)
leexy@buaa.edu.cn (Xiao-Yang Li)

Corresponding author*: Xiao-Yang Li

## Abstract

Accurate modeling of dependent degradation processes is essential for credible reliability assessment and remaining useful life (RUL) prediction in complex systems. Existing dependent degradation models usually describe dependence using correlation-based methods with symmetric characteristics. However, they ignore inherent causal directionality between degradation paths, which may lead to biased reliability evaluation and RUL predictions when physical causality exists. To address this issue, this paper proposes a causality-driven framework for modeling dependent degradation data. Firstly, univariate degradation models with multi-source uncertainties are established for each performance indicator based on the Wiener process. Then, the stable Peter-Clark algorithm is employed to uncover inherent causal relationships between degradation processes, and an uncertainty-aware neural network is employed to quantify causal effects considering uncertainties. Next, univariate degradation predictions and causal predictions are integrated within a Bayesian framework to construct a causally dependent degradation model, and the corresponding loss function is derived for model training. Finally, system reliability and RUL predictions are derived via Monte Carlo simulation. The proposed methodology is validated on the C-MAPSS dataset. Results show that considering inherent causal directionality between degradation processes helps eliminate physically unrealistic degradation behaviors, yielding more accurate degradation and RUL predictions than independent and correlation-based dependent degradation models.

Keywords: Dependent degradation modeling; multivariate degradation; causal discovery; reliability analysis; remaining useful life.

## Highlights

- Causal relationships are considered in dependent degradation modeling.
- Degradation predictions and causal predictions are fused in a Bayesian framework.
- Reliability and RUL predictions are derived based on the proposed model.
- Ignoring causality can lead to large deviations of degradation and RUL predictions.
- The proposed method offers greater accuracy than the correlation-based approaches.

# 1 Introduction

With the increasing complexity of industrial systems and the widespread deployment of sensors, prognostics and health management (PHM) has become an important basis for condition-based maintenance and risk-informed decision making [1]. Degradation modeling describes the performance degradation before failure and therefore plays a key role in reliability analysis and remaining useful life (RUL) prediction for highly reliably products. Most existing work on degradation modeling focuses on a single performance characteristic. In this line of research, stochastic processes, such as Wiener processes [2, 3], Gamma processes [4], inverse Gaussian processes [5] and Tweedie exponential dispersion processes [6], are widely adopted to describe the degradation of a scalar health indicator, and the RUL distributions are obtained through first hitting time analysis. Furthermore, recent studies incorporate memory effects [7, 8], change point [9] and physics-informed neural networks [10] into stochastic processes, which improves the accuracy and adaptability of degradation models in complex environments. However, these methods implicitly assume that the system performance can be sufficiently represented by a single parameter. This assumption is often not realistic for modern systems characterized by multiple performance parameters. For example, in aero engines, thrust, compressor efficiency and fuel flow rate jointly characterize engine performance, while in rotating machinery and electromechanical drives, output torque, efficiency and vibration amplitude are often considered simultaneously.

In this context, statistical multivariate degradation modeling has attracted growing interest [11-14], with various correlation structures being incorporated into degradation models. A notable branch of this research emphasizes capturing the correlations between a specific model parameter in different univariate degradation models, with the covariance matrix employed to quantify such correlations. For instance, Lu et al. [15] proposed a multivariate degradation model that accounts for correlations in degradation rates of different univariate degradation models. Wu et al. [16] considered the correlation of diffusion processes across univariate degradation models. Zheng et al. [17] proposed a degradation model that accounts for the correlation of initial performance. Asgari et al. [18, 19] investigated correlations in degradation rates and memory effects, respectively. In parallel, copula functions have been applied to characterize dependencies between probabilistic distributions derived from univariate degradation models. Sari et al. [20] posited that failures resulting from the degradation of two performance parameters are correlated, and utilized copula functions to model the correlation between first hitting time distributions (i.e., lifetime distributions). Extending this idea to multivariate degradation processes, Xu et al. [21] introduced vine copulas to capture the correlations among multiple lifetime distributions. Subsequently, Wen et al. [22] quantified the multiple sources of uncertainty involved in the model. Sun et al. [23] further extended the vine copula approach to accelerated degradation modeling. Meanwhile, some researchers have directly focused on the degradation process itself, examining the dependence among degradation increments or degradation values. Peng et al. [24] employed copula functions to characterize the dependence in the distributions of degradation increments for two degradation paths. Furthermore, Fang and Pan [25] extended this approach to multivariate degradation by considering five types of copula functions along with different vine copula structures. On the other hand, Rodríguez-Picón et al. [26] modeled the correlation between degradation values from two degradation processes using copula functions. In addition to the aforementioned methods, Wu et al. [27-29] offered a distinct perspective by suggesting that the degradation rate of a given performance parameter is influenced by the states of other parameters, with their interactions represented via a state-space correlation coefficient matrix. All these approaches have demonstrated effectiveness in degradation modeling and RUL prediction of systems with multiple performance indicators. Table 1 presents a summary of representative approaches for modeling dependent degradation processes.

Table 1 Summary of representative approaches for modeling dependent degradation processes.

| Category | Modeling strategy | Target | Work |
| --- | --- | --- | --- |
| Correlation of degradation model parameters | Covariance matrix | Initial performance | [17] |
| | | Degradation rate | [15, 18] |
| | | Diffusion processes | [16, 19] |
| Correlation of degradation model results | Copula function | Lifetime distributions | [21-23] |
| Correlation of degradation path characteristics | Copula function | Degradation increments | [24, 25] |
| | | Degradation values | [26] |
| Correlation of degradation interaction | State-space correlation coefficient matrix | Degradation state and degradation rate | [27-29] |

However, correlation-based dependence modeling is inherently symmetric and lacks explicit causal directionality between performance parameters. As a result, it cannot determine whether one parameter influences another. In many systems, distinct physical causal pathways exist. For example, in an aero engine, fuel flow (FF) [30] and exhaust gas temperature (EGT) [31] are two key performance indicators. During operation, an increase in FF leads to more fuel being injected into the combustor, which intensifies the combustion process, raises the gas temperature and consequently results in a higher EGT. Thus, FF exerts a causal influence on EGT. A purely correlation-based approach, however, treats this relationship as symmetric and does not distinguish the physically meaningful direction from FF to EGT. It may therefore support the misleading inference that a decrease in EGT implies an increased likelihood of FF reduction. In practice, FF is scheduled by the control system according to thrust demand and operational constraints, and it is not adjusted due to the variation of EGT. This example shows that uncovering the causal relationships between degradation processes is necessary to represent their degradation dependencies accurately and to avoid erroneous conclusions drawn from symmetric correlation measures.

In recent years, the importance of causality has begun to receive attention in PHM. Several works introduce causal graphs into deep learning architectures for multivariate time series, combining causal discovery with graph neural networks for fault detection [32-34] and fault diagnosis [35, 36]. These studies show that causal information can enhance the prediction accuracy and interpretability in multi-sensor PHM tasks. As for degradation-related applications, Zheng et al. [37] developed a spatial-temporal attention network based on causal graphs derived from degradation trajectories to enhance feature learning across multiple sensors, which was then applied to RUL estimation in complex systems. Nevertheless, they treat RUL as a supervised learning target and focus on black-box models. Explicit degradation path modeling and classical reliability quantities, such as reliability functions and first hitting time distributions, are not considered. In addition, causal structures are introduced at the feature level rather than at the level of stochastic degradation processes.

Therefore, there is still a gap between two parallel research directions. On one hand, statistical multivariate degradation modeling provides rigorous stochastic process-based tools for reliability analysis, but always describes dependence using symmetric correlation measures without explicit inherent causal relationships. Here, the inherent causal relationships refer to the direct physical or functional dependencies between performance parameters, where the variation of one parameter causally influences another. On the other hand, causality-aware deep learning approaches for RUL prediction use causal discovery to improve prediction performance, but they seldom construct a transparent dependent degradation model that can support degradation prediction and reliability analysis, and they do not fully exploit the physical interpretability of stochastic process-based degradation models that have been extensively studied in reliability engineering. Bridging this gap requires a framework that can explicitly determine causal relationships between degradation paths, and that can integrate these relationships with univariate degradation processes within a unified framework.

To bridge this gap, this study introduces a causality-driven framework for modeling dependent degradation in systems characterized by multiple performance indicators. The basic idea is to describe each performance parameter using an appropriate stochastic degradation process individually and to characterize their dependence through a learned causal graph and causal effect functions. Specifically, univariate degradation models considering multi-source uncertainties are first established for each performance parameter based on the Wiener process. Then, the inherent causal relationships between degradation processes are determined using the stable Peter-Clark (Stable-PC) causal discovery technique. To quantify causal effects, an uncertainty-aware neural network is introduced, which maps parent degradation states to child states and outputs predictive distributions so that uncertainty in the causal effects can be incorporated. Finally, the causal influence model is integrated with univariate degradation models within a Bayesian framework, and the corresponding loss function is derived for model training. On this basis, system reliability and RUL are evaluated.

The main contributions of this work are summarized as follows.

- A causality-driven framework for modeling dependent degradation data is developed, incorporating causal influence modeling and stochastic process-based degradation modeling.
- An uncertainty-aware neural network is employed to provide probabilistic quantification of causal effects between degradation processes, and univariate degradation models with multi-source uncertainties are integrated with the causal effect model within a Bayesian framework.
- The effectiveness of the proposed model is demonstrated through the C-MAPSS dataset, where the method is shown to improve the accuracy of degradation and RUL predictions compared with correlation-based dependent degradation models.

The organization of the paper is as follows. Firstly, Section 2 gives the proposed method, including the causally dependent degradation modeling framework and procedures for reliability analysis and RUL predictions. Then, in Section 3, the superiority of the proposed methodology are verified by C-MAPSS dataset. Finally, we conclude our work in Section 4.

# 2 Proposed method

The proposed methodology is summarized in Fig. 1. Based on the observed dependent degradation data, the approach proceeds along two directions. First, a univariate degradation model is constructed for each performance parameter, capturing both the deterministic degradation trend and multi-source uncertainties. Second, causal relationships between the degradation paths are identified, and the corresponding causal influences are modeled using an uncertainty-aware neural network. These two components are then integrated within a Bayesian framework, where prior information from the univariate degradation model is combined with causality information from the causal influence to construct a causally dependent degradation model. The corresponding model training method is provided. Finally, the proposed model is used to perform reliability analysis and RUL prediction for dependent degradation processes. For clarity, the main notations used in this paper are summarized in Table 2.

Table 2 Summary of main notations.

| Symbol | Description |
|---|---|
| $B_s$ | Number of Monte Carlo samples |
| $B(t)$ | Wiener process |
| $K$ | Number of performance parameters |
| $M(t)$ | System performance margin |
| $N_T$ | Sample size used in causal discovery |
| $\mathrm{Pa}(Y_k)$ | Parent set of the $k^{\mathrm{th}}$ performance parameter in the causal graph |
| $R(t)$ | System reliability |
| $T_{\mathrm{RUL}}$ | Remaining useful life |
| $Y_0$ | Initial performance value |
| $Y_k$ | The $k^{\mathrm{th}}$ performance parameter |
| $Y_{k,\mathrm{th}}$ | Failure threshold of the $k^{\mathrm{th}}$ performance parameter |
| $a$ | Degradation rate parameter |
| $f_{\mu^{(k)}}$ | Neural network for calculating the mean of causal impact for $Y_k$ |
| $f_{\sigma^{(k)}}$ | Neural network for calculating the standard deviation of causal impact for $Y_k$ |
| $i, j, k$ | Indices for unit, time, and variable |
| $m_i$ | Number of observations for the $i^{\mathrm{th}}$ unit |
| $n$ | Number of units |
| $t_{ij}$ | Observation time of the $i^{\mathrm{th}}$ unit at index $j$ |
| $y_{ij}^{(k)}$ | Observation of $Y_k$ for the $i^{\mathrm{th}}$ unit at time $j$ |
| $\mathbf{y}_{ij}^{\mathrm{Pa}(Y_k)}$ | Observation vector of parent variables of $Y_k$ |
| $\Delta y_{ij}^{(k)}$ | Degradation increment of $Y_k$ for the $i^{\mathrm{th}}$ unit at index $j$ |
| $\mathcal{D}$ | Dataset used for causal discovery |
| $\mathbf{\Xi}_i$ | Covariance matrix of observations for unit $i$ |
| $\Psi(t)$ | Time-scale function |
| $\beta$ | Parameter of the time-scale function |
| $\varepsilon$ | Measurement error |
| $\boldsymbol{\theta}$ | Parameter vector of the degradation model |
| $\sigma$ | Diffusion coefficient |

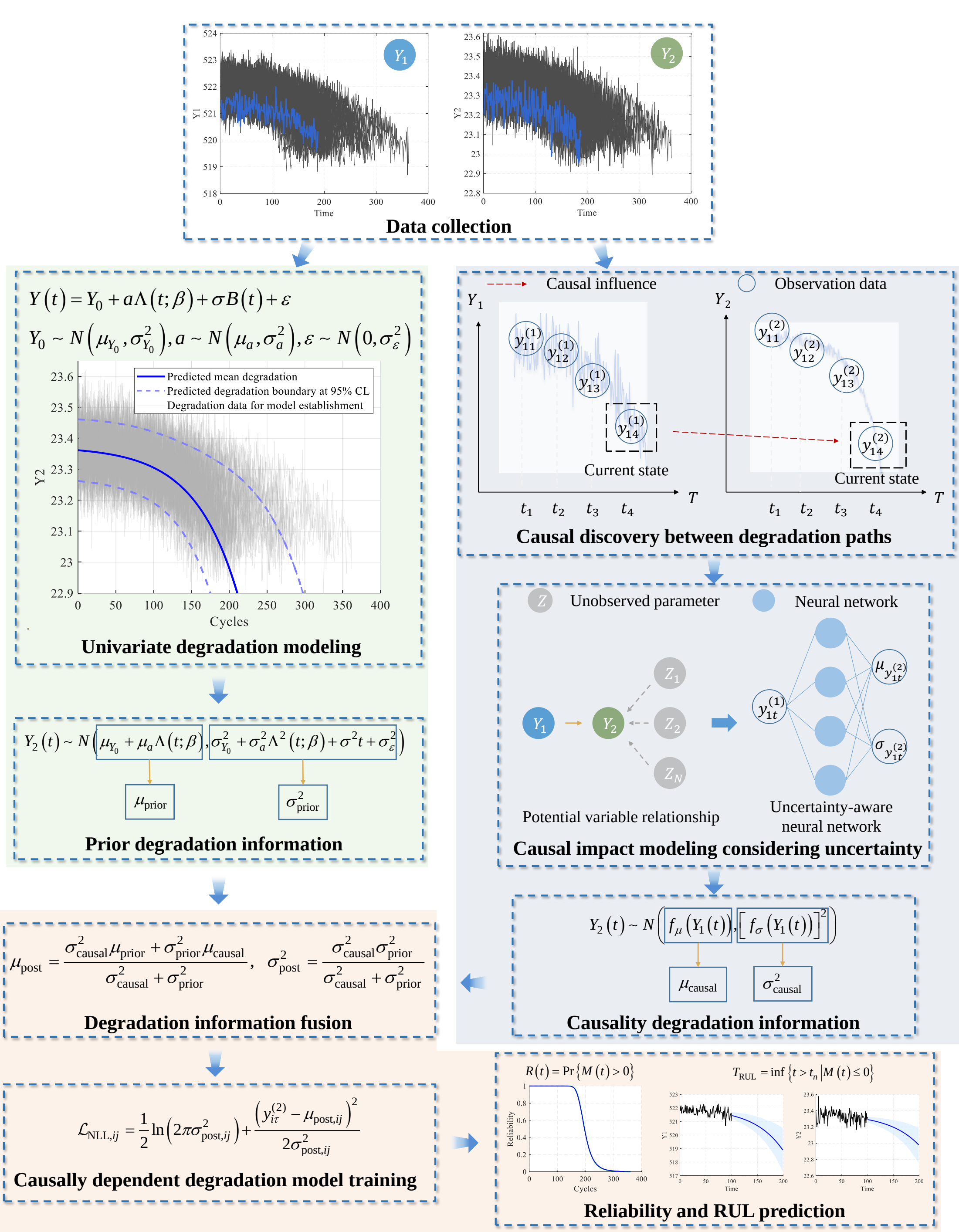

Fig. 1 The proposed research methodology.

### 2.1 Univariate degradation modeling

In this section, a univariate degradation model is established for individual degradation paths, along with the statistical analysis method to estimate the unknown model parameters from the observed data.

#### 2.1.1 Univariate degradation model establishment

Under specific operating conditions, the degradation process of the system is primarily governed by a deterministic degradation pattern, which can be described by [38, 39]:

$$Y(t) = Y_0 + a\Psi(t;\beta), \tag{1}$$

where $Y$ represents the performance parameter of the system; $Y_0$ is the initial value of the performance without considering degradation; $t$ is the degradation time; $a$ is the unknown parameter represents the degradation rate; and $\Psi(\cdot)$ is the time-scale function with unknown parameters $\beta$. Common time-scale functions used in the literature include linear, power ($t^{\beta}$) and exponential forms ($e^{\beta t}-1$) [40]. The selection of the time-scale function is typically guided by prior knowledge or model selection techniques.

In addition to the deterministic degradation trend, the degradation process of the system is also influenced by multiple sources of uncertainty. Firstly, imperfections in the manufacturing process introduce variability in both the initial performance values and degradation rates across individual units [41]. Furthermore, the uncertainty in the degradation process increases over time, reflecting temporal variability [42]. Additionally, noise during the observation results in measurement errors [43]. These sources of uncertainty have been extensively studied in the literature, and mature quantification methods are available. Without loss of generality, this study assumes that uncertainties in the initial performance, degradation rate (also known as random effects) and measurement error follow normal distributions, which is consistent with most existing literature. For temporal variability, it is modeled using a Wiener process. Compared to Gamma and inverse Gaussian processes, which require strict monotonicity for method validity, the Wiener process can describe degradation data with fluctuations and offers greater generality [10]. Even for degradation processes that are inherently monotonic, the Wiener process can still be used to model the overall degradation trend when the drift coefficient dominates [44, 45].

Based on the above analysis, the degradation model in Eq. (1) can be extended by considering multi-source uncertainties as:

$$\begin{gathered} Y(t) = Y_0 + a\Psi(t;\beta) + \sigma B(t) + \varepsilon, \\ Y_0 \sim N\left(\mu_{Y_0}, \sigma_{Y_0}^2\right), a \sim N\left(\mu_a, \sigma_a^2\right), \varepsilon \sim N\left(0, \sigma_\varepsilon^2\right), \end{gathered} \tag{2}$$

where $\mu_i$ and $\sigma_i$ represent the mean and standard deviation of the random variable $i$, respectively; $\sigma$ is the diffusion coefficient; $\varepsilon$ is the measurement error; $B(t)$ is the Wiener process; and $N$ denotes the normal distribution.

#### 2.1.2 Statistical analysis method for univariate degradation models

For the degradation model described by Eq. (2), the unknown parameters in the model are $\mathbf{\theta} = \left[\mu_{Y_0}, \sigma_{Y_0}, \mu_a, \sigma_a, \beta, \sigma, \sigma_\varepsilon\right]$. These parameters need to be determined using the observed degradation data. Let the $j^{\text{th}}$ performance observation for the $i^{\text{th}}$ unit be denoted as $y_{ij}$, which is recorded at time $t_{ij}$. Here, $i$ = 1, 2,..., $n$ indexes the units, and $j$ = 1, 2,..., $m_i$ indexes the measurements for unit $i$, where $n$ is the total number of units and $m_i$ is the number of observations for unit $i$. Hereby, we denote $\mathbf{y}_i = \left[y_{i1}, y_{i2}, \ldots, y_{im_i}\right]^{\mathrm{T}}$ and $\boldsymbol{\Psi}_i = \left[\Psi(t_{i1};\beta), \Psi(t_{i2};\beta), \ldots, \Psi(t_{im_i};\beta)\right]^{\mathrm{T}}$. Then, according to the property of the Wiener process, $\boldsymbol{y}_i$ follows a multivariate normal probability distribution given by:

$$\mathbf{y}_i \sim N\left(\mu_{Y_0} + \mu_a \boldsymbol{\Psi}_i, \boldsymbol{\Xi}_i\right), \tag{3}$$

where $\boldsymbol{\Xi}_i$ is a covariance matrix of dimension $m_i \times m_i$, with its $(u, v)^{\text{th}}$ element calculated by:

$$\left(\boldsymbol{\Xi}_i\right)_{uv} = \sigma_{Y_0}^2 + \sigma_a^2 \Psi(t_{iu};\beta)\Psi(t_{iv};\beta) + \sigma^2 \min(t_{iu}, t_{iv}) + \delta_{uv}\sigma_\varepsilon^2, \tag{4}$$

where $\delta_{uv}$ denotes the Kronecker delta function, which equals 1 when $u = v$ and 0 otherwise.

Subsequently, the log-likelihood function of Eq. (2) given $\mathbf{y}_i$ can be derived as:

$$\ln L_i\left(\boldsymbol{\theta}|\mathbf{y}_i\right) = -\frac{1}{2}\left[m_i \ln(2\pi) + \ln|\boldsymbol{\Xi}_i| + \left(\mathbf{y}_i - \mu_{Y_0} - \mu_a \boldsymbol{\Psi}_i\right)^{\mathrm{T}} \boldsymbol{\Xi}_i^{-1} \left(\mathbf{y}_i - \mu_{Y_0} - \mu_a \boldsymbol{\Psi}_i\right)\right]. \tag{5}$$

Based on Eq. (5), when given all the observations $\mathbf{y} = \left[\mathbf{y}_1, \mathbf{y}_2, \ldots, \mathbf{y}_n\right]$, the log-likelihood function can be expressed as:

$$\ln L_{\text{all}}\left(\boldsymbol{\theta}|\mathbf{y}\right) = \sum_{i=1}^{n} \ln L_{li}\left(\boldsymbol{\theta}|\mathbf{y}_i\right). \tag{6}$$

By maximizing the log-likelihood function described in Eq. (6), the estimates of $\boldsymbol{\theta}$ can be obtained as:

$$\hat{\boldsymbol{\theta}} = \arg\max_{\boldsymbol{\theta}} \ln L_{\text{all}}\left(\boldsymbol{\theta}|\mathbf{y}\right). \tag{7}$$

Due to the non-convex and nonlinear characteristics of Eq. (6), conventional gradient-based methods may converge to local optima. Therefore, an improved variant of the meta-heuristic rime algorithm, termed TERIME [46], is adopted due to its effectiveness in handling high-dimensional optimization tasks. The source code of the TERIME algorithm is provided in https://github.com/dirge1/TERIME. Following the recommendations in [46, 47], the population size $P_s$ is set to 20, and the maximum number of iterations $T_{\max}$ is set to 3000. The parameter estimation procedure is outlined in Algorithm 1.

**Algorithm 1**: Statistical analysis method employing TERIME to estimate the unknown parameters based on observed degradation data.

**Input**:
1. Population size $P_s$ and maximum iteration count $T_{\max}$ for TERIME.
2. Parameter bounds (lower and upper limits).
3. Degradation observations $\boldsymbol{y}$ and their corresponding measurement time.

**Output**:
1. Estimates of unknown parameters in the univariate degradation model $\hat{\boldsymbol{\theta}}$.
2. Maximum value of the log-likelihood function.

**Objective function**:
1. Maximize the log-likelihood function given in Eq. (6).

**Procedure**:
1. Randomly generate $P_s$ candidate solutions within parameter bounds.
2. For each candidate, compute the objective function via Eq. (6).
3. Update candidate solutions according to the rules of TERIME [46].
4. Stop after $T_{\max}$ iterations.
5. Output the candidate with the highest log-likelihood function as $\hat{\boldsymbol{\theta}}$.

### 2.2 Dependent degradation modeling considering inherent causal relationships

In this section, a causal discovery method is introduced to determine causal dependencies among degradation paths. Then, a causally dependent degradation model is established and a corresponding model training method is developed.

#### 2.2.1 Causal discovery among degradation paths

In engineering systems, causal relationships between degradation paths may be obtained from different sources. When explicit physical mechanisms are available, causal relationships can be specified according to domain knowledge. However, for complex systems with multiple interacting performance indicators, such relationships are often only partially known or difficult to formulate explicitly. In such cases, data-driven causal discovery provides an effective tool for identifying causal dependencies from observed degradation data.

Accordingly, this study combines physical knowledge and data-driven causal discovery to construct the causal structure among degradation paths. The causal discovery method is used to identify causal dependencies supported by observational data, while domain knowledge is incorporated when the direction of an edge cannot be uniquely determined from data alone. This strategy helps ensure that the resulting causal graph is both data-consistent and physically meaningful.

Let $y_{ij}^{(k)}$ denote the $j^{\text{th}}$ observation of the $k^{\text{th}}$ performance parameter for the $i^{\text{th}}$ unit, measured at time $t_{ij}$, where $k$ = 1, 2,..., $K$ indexes the monitored parameters and $K$ is the total number of performance parameters. For causal structure learning, direct use of raw degradation data can lead to spurious dependencies caused by their non-stationary trends. To mitigate this, the raw degradation paths are transformed into degradation increments following our previous study [48], given by $\Delta y_{ij}^{(k)} = y_{ij}^{(k)} - y_{i,j-1}^{(k)}$, $j \geq 2$. Subsequently, at each measurement time $t_{ij}$, the contemporaneous increment vector is defined as:

$$\Delta \mathbf{y}_{ij} = \left(\Delta y_{ij}^{(1)}, ..., \Delta y_{ij}^{(M)}\right)^{\mathrm{T}}. \tag{8}$$

Accordingly, the dataset $\mathcal{D}$ for causal discovery across all units and measurement times is given by:

$$\mathcal{D} = \left\{\Delta \mathbf{y}_{ij} : i = 1, ..., n;\ j = 2, ..., m_i\right\}, \tag{9}$$

with total sample size $N_T = \sum_{i=1}^{n} (m_i - 1)$.

Considering the physical characteristics of degradation processes in engineering systems, this work introduces the following two assumptions regarding causal relationships between degradation paths [48].

**Assumption 1** (Contemporaneous causality): If a causal relationship exists from degradation path $Y_1$ to degradation path $Y_2$, it implies that the current performance state of $Y_1$ causally affects the current performance state of $Y_2$, whereas the historical states of $Y_1$ have no causal influence on $Y_2$, as illustrated in Fig. 2. This assumption holds because system performance is typically observed under steady-state conditions, where causal relationships reflect underlying physical principles between performance parameters.

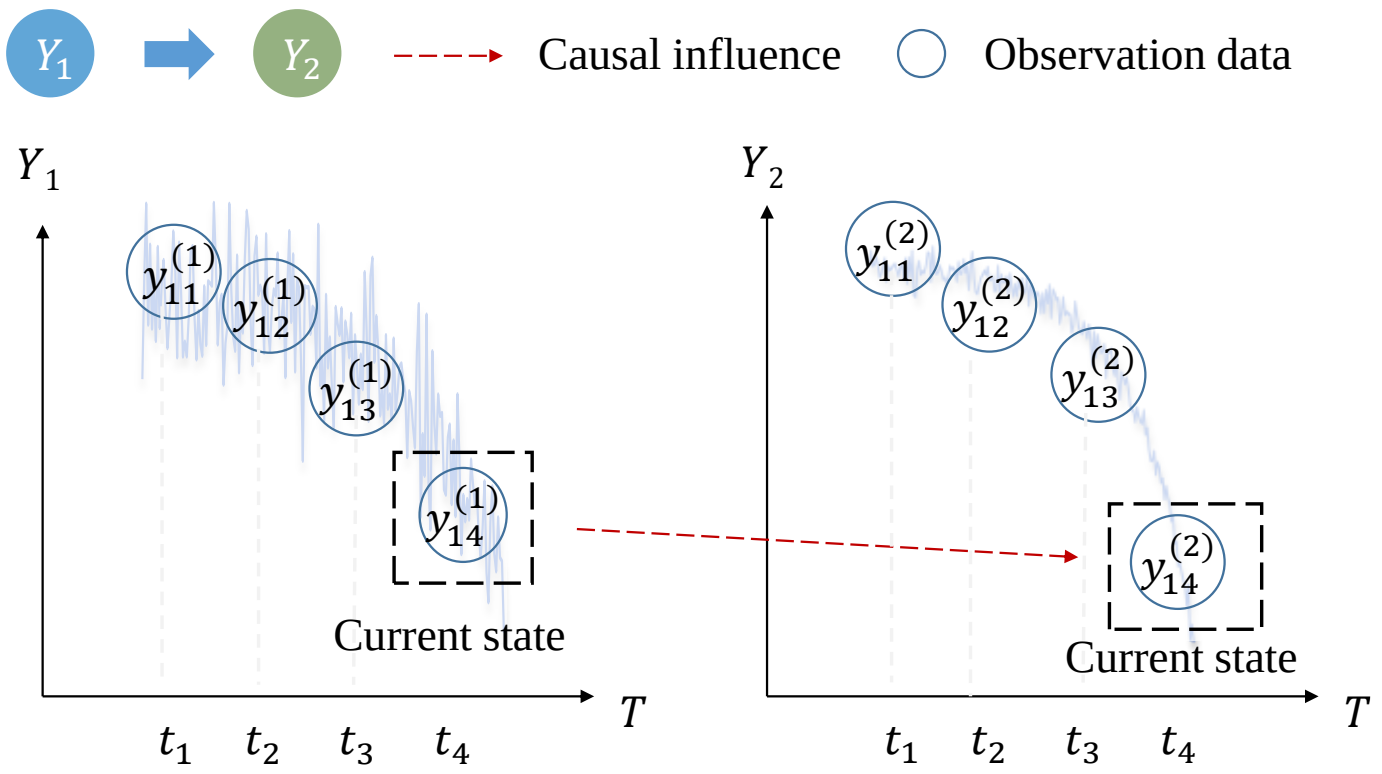

Fig. 2 Schematic of the causal relationship from degradation path $Y_1$ to degradation path $Y_2$.

**Assumption 2** (Stability of causal dependencies): Prior to system failure, the causal dependencies between degradation paths under steady-state conditions remain stable and invariant over time. This assumption holds because the underlying physical principles governing the dependencies of degradation paths remain unchanged.

These assumptions are consistent with many engineering systems, where performance indicators are directly determined by the instantaneous states of internal components. For instance, in analog circuits, performance indicators such as center frequency and gain are determined by the current values of resistors and capacitors through stable physical relationships. In this case, the observed dependencies between performance parameters are inherently contemporaneous and can be considered time-invariant over the degradation phase. Nevertheless, it should be noted that these assumptions may not hold in systems with strong dynamic behaviors, such as time-delay effects or feedback mechanisms across time. In such cases, temporal causal discovery methods or dynamic causal models would be more appropriate [49]. Therefore, the proposed framework is mainly applicable to degradation processes measured under steady-state conditions, where the causal relationships can be reasonably approximated as contemporaneous and stable over time.

In our previous study [48], we compared the effectiveness of four different causal discovery methods in identifying causal relationships between degradation paths, including stable Peter-Clark (Stable-PC) [50], Greedy Equivalence Search (GES) [51], Direct Linear Non-Gaussian Acyclic Model (Direct-LiNGAM) [52], and Non-combinatorial Optimization via Trace Exponential and Augmented Lagrangian for Structure Learning (NOTEARS) [53, 54]. The results indicated that Stable-PC provides more consistent and robust identification of causal relationships. Therefore, Stable-PC is adopted in this study to detect causal relationships among degradation paths.

Let $V = \{v_1,...,v_K\}$ be the vertex set, where $v_k$ represents the $k^{\text{th}}$ performance parameter. The procedure of Stable-PC consists of three main steps:

**Step 1. Skeleton learning**

The algorithm starts with a complete undirected graph over the variable set $V$ and sets the conditioning set size $l = 0$. At each level $l$, the algorithm freezes the adjacency sets at the beginning to ensure order-independence. For each existing edge ($v_a$, $v_b$), the conditioning set $S$ can be given as:

$$S \subseteq \text{Adj}(v_a) \setminus \{v_b\}, |S| = l, \tag{10}$$

where $\mathrm{Adj}(v_a)$ represents the adjacency sets of $v_a$. Then, we test the null hypothesis that $v_a$ and $v_b$ are conditionally independent given $S$. Under a Gaussian approximation, this is assessed by the partial correlation as:

$$\rho_{ab|S} = -\frac{\Sigma_{ab}^{-1}}{\sqrt{\Sigma_{aa}^{-1}\Sigma_{bb}^{-1}}}, \tag{11}$$

where $\Sigma_{ab}^{-1}$ denotes the $(a, b)^{\text{th}}$ entry of the precision matrix for variables $S \cup \{v_a, v_b\}$, and $\Sigma_{aa}^{-1}$ and $\Sigma_{bb}^{-1}$ are the corresponding diagonal entries. The null hypothesis $H_0 : \rho_{ab|S} = 0$ is tested using the Fisher z-statistic given by:

$$z = \frac{1}{2}\ln\left(\frac{1+\rho_{ab|S}}{1-\rho_{ab|S}}\right)\sqrt{N_T - |S| - 3}, \tag{12}$$

which converges in the standard normal distribution as $N_s \to \infty$ under $H_0$. Using a significance level $\alpha$ (typically 0.05), if the test fails to reject $H_0$, the variables are considered conditionally independent, and the edge ($v_a$, $v_b$) is marked for removal. It is worth noting that the effective sample size for causal discovery is obtained by aggregating degradation increments across all units and observation times, as given in Eq. (9), resulting in a sufficiently large dataset to support the asymptotic validity of the Fisher z-test.

After all candidate sets of size $l$ are tested for all edges, the edges marked as independent are deleted simultaneously. The conditioning set size $l$ is then increased by one, and the procedure repeats until no further tests are possible.

At the end of this step, we obtain an undirected skeleton and a collection of separating sets $S(a,b)$ for each pair of variables ($v_a$, $v_b$), where $S(a,b)$ contains all subsets that render $v_a$ and $v_b$ conditionally independent.

**Step 2. Orientation of V-structures**

For every unshielded triplet ($v_a$, $v_b$, $v_c$), where $v_a$ and $v_c$ are not adjacent but both connected to $v_b$, some orientations can be deduced based on the property of V-structures: if $v_b \notin S_c$ for all $S_c \in S(a,c)$, it can be concluded that $v_b$ is a collider and the triple is oriented as $v_a \rightarrow v_b \leftarrow v_c$. On the other hand, if $v_b \in S_c$ for at least one separating set $S_c \in S(a,c)$, it can be concluded that $v_b$ is not a collider and the triple remains unoriented.

**Step 3. Orientation Propagation**

After orienting all unshielded triples, the algorithm applies the Meek rules [55] to infer additional edge directions while preventing cycles and spurious v-structures. These rules are applied iteratively until no further orientations are possible:

Rule 1: If $v_a \rightarrow v_b$ , $v_b \sim v_c$ (adjacent but undirected), and $v_a \nsim v_c$ (not adjacent), orient $v_b \rightarrow v_c$ .

Rule 2: If $v_a \sim v_b$ and there exists a directed path $v_a \rightarrow v_c \rightarrow v_b$, orient $v_a \rightarrow v_b$.

Rule 3: If $v_a \sim v_b$ and there exist two chains $v_a \sim v_c \rightarrow v_b$ and $v_a \sim v_d \rightarrow v_b$ with $v_c \nsim v_d$, orient $v_a \rightarrow v_b$.

Rule 4: If $v_a - v_b$ and there exist two chains $v_a \sim v_d \rightarrow v_b$ and $v_a \sim v_c \rightarrow v_d$ with $v_b \nsim v_c$, orient $v_a \rightarrow v_b$.

Fig. 3 illustrates these orientation rules with concrete examples.

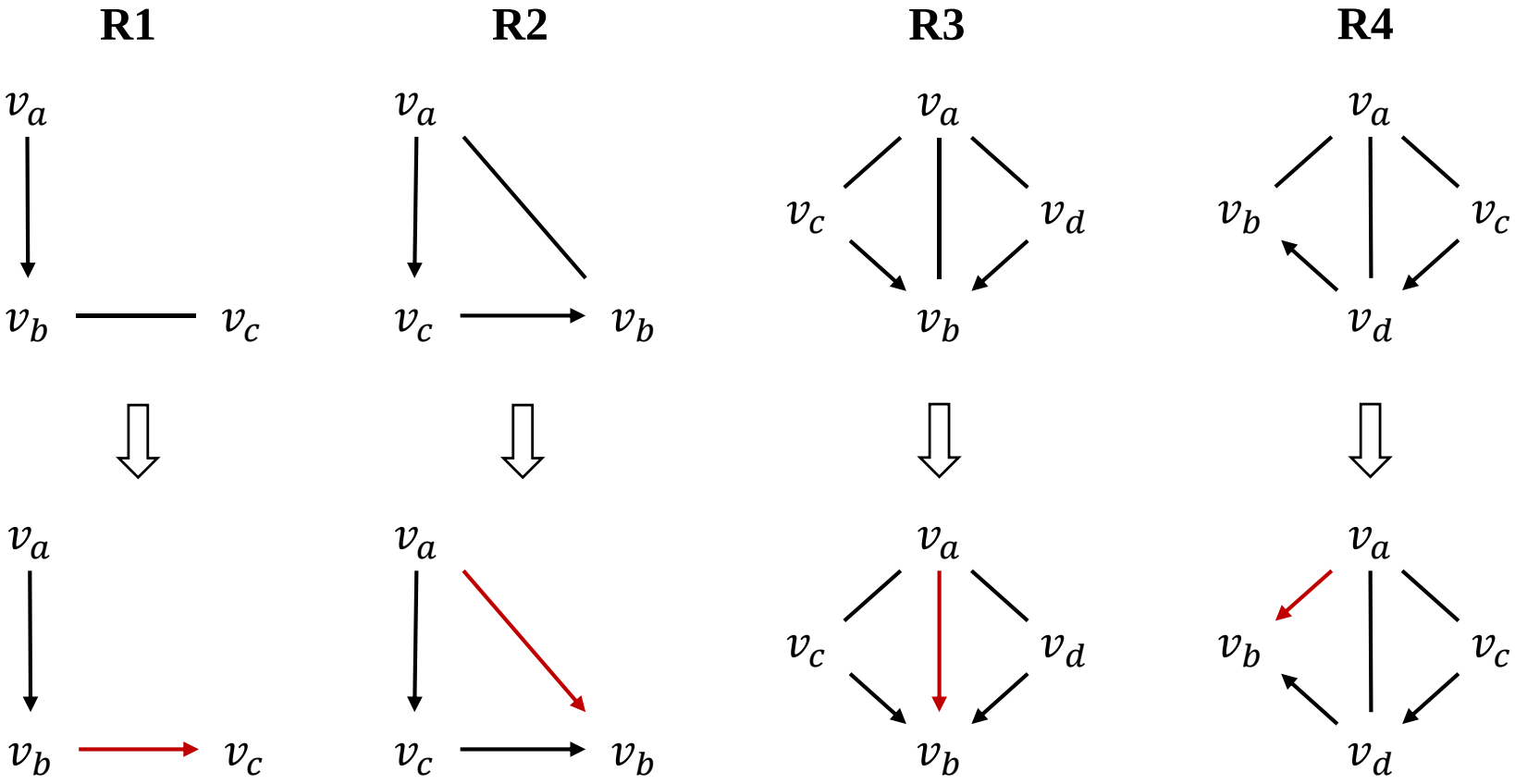

Fig. 3 The orientation rules from Meek. If the top graph appears as an induced subgraph, orient the red edge as shown in the corresponding bottom graph.

**Step 4. Orientation using domain knowledge**

After the above three steps, the algorithm produces a completed partially directed acyclic graph representing the Markov equivalence class of causal structures supported by the observed data. In general, the resulting graph may contain undirected edges when the available conditional independence information is insufficient to determine their orientation. In the special case $K = 2$, the skeleton reduces to a single undirected edge, as no third variable exists for conditioning. When undirected edges exist in the resulting causal graph, domain knowledge is incorporated to determine the directions of the remaining edges, ensuring that the inferred causal directions are consistent with the underlying system mechanisms. When the direction remains ambiguous, predictive performance can be used as a criterion, as the correct causal direction typically yields better performance (see Section 3.7.2). In summary, this step complements the data-driven causal discovery process and results in a fully directed causal graph for subsequent modeling.

The detailed procedures for causal discovery among degradation paths using Stable-PC algorithm are summarized in Algorithm 2.

---

**Algorithm 2**: Causal structure learning for degradation paths using Stable-PC algorithm.

---

**Input**:

1. Dataset $\mathcal{D} = \left\{\Delta\mathbf{y}_{ij} : i = 1,\ldots,n;\ j = 2,\ldots,m_i\right\}$ with total sample size $N_T$.
2. Variable set $V = \left\{v_1,\ldots,v_K\right\}$.
3. Significance level $\alpha$ (typically $\alpha = 0.05$).

**Output**:

1. Completed partially directed acyclic graph for describing causal relationship among degradation paths.

**Procedure**:

1. Initialization:

1.1 Construct a complete undirected graph $G^{(0)}$ over $V$.

1.2 Set conditioning set size $l \leftarrow 0$.

---

1.3 Initialize separating sets: $S(a,b) \leftarrow \varnothing$ for all existing edge ($v_a$, $v_b$).
2. Skeleton learning:
2.1 **while** there exists a node $v_a$ such that $\mathrm{Adj}(v_a) - 1 \geq l$ **do**:
2.2 **for** each existing edge ($v_a$, $v_b$) in $G^{(l)}$ **do**:
2.3 **for** each subset $S \subseteq \mathrm{Adj}(v_a) \setminus \{v_b\}$ with $|S| = l$ **do**:
2.4 Compute partial correlation via Eq. (11).
2.5 Compute Fisher-z statistic via Eq. (12).
2.6 **if** $|z| \leq z_{1-\alpha/2}$ **then**:
2.7 Mark edge ($v_a$, $v_b$) for removal.
2.8 Update separating sets: $S(a,b) \leftarrow S(a,b) \cup \{S\}$.
2.9 **end if**
2.10 **end for**
2.11 **end for**
2.12 Delete all marked edges simultaneously.
2.13 Update level: $l \leftarrow l + 1$.
2.14 **end while**
3. Orientation of V-structures:
3.1 **for** each unshielded triple ($v_a$, $v_b$, $v_c$) with $v_a \sim v_b$, $v_b \sim v_c$ and $v_a \nsim v_c$ **do**:
3.2 **if** $v_b \notin S_c$ for all $S_c \in S(a,c)$ **then**:
3.3 Orient collider: $v_a \rightarrow v_b \leftarrow v_c$.
3.4 **end if**
3.5 **end for**
4. Orientation propagation:
4.1 **repeat**
4.2 R1: **if** $v_a \rightarrow v_b$ , $v_b \sim v_c$ and $v_a \nsim v_c$ **then** $v_b \rightarrow v_c$
4.3 R2: **if** $v_a \sim v_b$ and $v_a \rightarrow v_c \rightarrow v_b$ **then** $v_a \rightarrow v_b$
4.4 R3: **if** $v_a \sim v_b$, $v_a \sim v_c \rightarrow v_b$, $v_a \sim v_d \rightarrow v_b$ and $v_c \nsim v_d$ **then** $v_a \rightarrow v_b$
4.5 R4: **if** $v_a - v_b$, $v_a \sim v_d \rightarrow v_b$, $v_a \sim v_c \rightarrow v_d$ and $v_b \nsim v_c$ **then** $v_a \rightarrow v_b$
4.6 **until** no more edges can be oriented.
5. Orientation using domain knowledge:
5.1 Detect edges that remain undirected after above steps.
5.2 Determine their directions based on prior knowledge of physical or functional dependencies.
5.3 Verify that the graph remains acyclic and consistent with physical laws.

### 2.2.2 Causally dependent degradation modeling

Once the causal relationships among degradation trajectories are identified, the modeling strategy can be formulated in a general form for multiple performance parameters. Let $\{Y_1, Y_2, \ldots, Y_K\}$ denote the set of performance indicators, and let $\mathrm{Pa}(Y_k)$ denote the set of parent variables of $Y_k$ in the inferred causal graph. For variables without causal parents, i.e., $\mathrm{Pa}(Y_k) = \varnothing$, the degradation process can be modeled independently using the univariate degradation models described in Section 2.1, and reliability analysis or RUL prediction can be conducted accordingly. For variables with causal dependencies, i.e., $\mathrm{Pa}(Y_k) \neq \varnothing$, the influence from parent variables should be explicitly modeled to describe the underlying physical relationships. It should be noted that the presence of causal effects does not imply that the target variable is fully determined by its parent variables, as it may also be affected by noise and other unobserved factors. To this end, an uncertainty-aware neural network is employed to model the conditional distribution of $Y_k$ given its parent variables $\mathrm{Pa}(Y_k)$ at the same observation time. The network outputs both the conditional mean and variance, enabling joint modeling of the expected value and predictive uncertainty. Specifically, the predictive distribution of $y_{ij}^{(k)}$ given $\mathbf{y}_{ij}^{\mathrm{Pa}(Y_k)}$ at time index $j$ for the $i^{\text{th}}$ unit is described by:

$$y_{ij}^{(k)} \left| \mathbf{y}_{ij}^{\mathrm{Pa}(Y_k)} \sim N\left( \mu_{ij}^{(k)}, \left[ \sigma_{ij}^{(k)} \right]^2 \right), \right. \tag{13}$$

where

$$\begin{aligned} \mu_{ij}^{(k)} &= f_{\mu^{(k)}}\left( \mathbf{y}_{ij}^{\mathrm{Pa}(Y_k)} \right), \\ \ln \sigma_{ij}^{(k)} &= f_{\sigma^{(k)}}\left( \mathbf{y}_{ij}^{\mathrm{Pa}(Y_k)} \right), \end{aligned} \tag{14}$$

$f_{\mu^{(k)}}$ and $f_{\sigma^{(k)}}$ are neural networks for calculating the predicted mean and standard deviation of the $k^{\text{th}}$ parameter, respectively; and $\mathbf{y}_{ij}^{\mathrm{Pa}(Y_k)}$ represents the corresponding vector of parent variables at time index $j$ for the $i^{\text{th}}$ unit.

Compared with parametric models that require predefined functional forms, the causal impacts between degradation processes are often nonlinear and difficult to specify explicitly. Therefore, a neural network is adopted as a flexible function approximator to learn these effects directly from data, avoiding restrictive structural assumptions. This design is consistent with recent advances in causal modeling, where neural networks are used to represent causal effects under given causal graphs [56-58]. Importantly, the interpretability of the proposed framework is primarily ensured by the causal graph, which explicitly encodes directional dependencies between degradation processes. Besides, it should be noted that the proposed framework is not limited to neural network-based implementations. In practical applications, deterministic models incorporating causal relationships are widely used, where dependencies between variables are governed by known physical laws or empirical relationships. In such cases, the neural network component can be replaced by physics-based or deterministic formulations. Moreover, when partial prior knowledge exists, knowledge-informed neural networks can be employed to integrate physical insights with data-driven modeling.

Once the causal effect is quantified, the degradation state of $Y_k$ at a specific time $t_{ij}$ is informed by two distinct sources of information. The first is the prediction from the univariate degradation model given by Eq. (2), which describes the overall degradation trend with uncertainties for the overall population. For a specific unit, its degradation state is expected to lie within the univariate degradation prediction for the population, and thus the prediction of the univariate degradation model can be regarded as the prior information of $Y_k$. According to Eq. (2), the prior mean and variance can be derived as:

$$\begin{aligned} \mu_{\mathrm{prior},ij}^{(k)} &= \mu_{Y_0}^{(k)} + \mu_a^{(k)} \Psi\left( t_{ij}; \beta^{(k)} \right), \\ \left[ \sigma_{\mathrm{prior},ij}^{(k)} \right]^2 &= \left[ \sigma_{Y_0}^{(k)} \right]^2 + \left[ \sigma_a^{(k)} \right]^2 \Psi^2\left( t_{ij}; \beta^{(k)} \right) + \left[ \sigma^{(k)} \right]^2 t_{ij}. \end{aligned} \tag{15}$$

where $i^{(k)}$ represents the parameter $i$ in the univariate degradation model of the $k^{\text{th}}$ performance parameter.

The second source is the prediction based on the causal effect conditioned on the contemporaneous observation of $\mathrm{Pa}(Y_k)$ given by Eq. (13). Since this prediction incorporates the causal influence of $\mathrm{Pa}(Y_k)$, it reflects unit-specific information and can therefore be regarded as the causality information of $Y_k$. Based on Eq. (13), the corresponding mean and variance of the causality distribution are:

$$\mu_{\text{causal},ij}^{(k)} = \mu_{ij}^{(k)},$$
$$\left[\sigma_{\text{causal},ij}^{(k)}\right]^2 = \left[\sigma_{ij}^{(k)}\right]^2. \tag{16}$$

The above two sources give complementary aspects of the state of $Y_k$: the prior information conveys the long-term, population-level degradation trend information with uncertainty, whereas the causality information captures contemporaneous, unit-specific information driven by causal influence from $\text{Pa}(Y_k)$. Relying solely on the prior overlooks the causal dependency between degradation paths for an individual unit, while relying solely on the causality distribution disregards the underlying degradation pattern and may suffer from instability in the presence of observational noise and unobserved parameters. Therefore, an integration of these two sources is conducted to leverage the strengths of both**.** Within a Bayesian framework, the two sources are coherently fused to obtain the mean and variance of the posterior predictive distribution as:

$$\mu_{\text{post},ij}^{(k)} = \frac{\left[\sigma_{\text{causal},ij}^{(k)}\right]^2 \mu_{\text{prior},ij}^{(k)} + \left[\sigma_{\text{prior},ij}^{(k)}\right]^2 \mu_{\text{causal},ij}^{(k)}}{\left[\sigma_{\text{causal},ij}^{(k)}\right]^2 + \left[\sigma_{\text{prior},ij}^{(k)}\right]^2},$$
$$\left[\sigma_{\text{post},ij}^{(k)}\right]^2 = \frac{\left[\sigma_{\text{causal},ij}^{(k)}\right]^2 \left[\sigma_{\text{prior},ij}^{(k)}\right]^2}{\left[\sigma_{\text{causal},ij}^{(k)}\right]^2 + \left[\sigma_{\text{prior},ij}^{(k)}\right]^2}. \tag{17}$$

Equation (17) arises from Bayesian updating with Gaussian prior and likelihood, where the posterior mean is a precision-weighted average of the two means, and the posterior variance gives more weight to the source with smaller uncertainty. This fusion ensures that the posterior prediction reflects the global degradation trend and the contemporaneous causal effect simultaneously, thereby achieving causally dependent degradation modeling and prediction.

Notably, although the proposed framework can be extended to multiple performance parameters, its practical applicability depends on the scale of the system. As the number of variables increases, the complexity of causal discovery and the modeling of causal dependencies grow significantly. Therefore, the method is most suitable for systems with a moderate number of performance parameters, while additional strategies would be required for large-scale systems.

### 2.2.3 Model training method

In the causally dependent degradation model defined by Eqs. (15)-(17), the prior distribution parameters in Eq. (15) are determined using the statistical estimation procedure in Section 2.1.2, while the causality distribution parameters in Eq. (16) are learned by training the neural networks $f_\mu(\cdot)$ and $f_\sigma(\cdot)$. For each observation of $Y_k$ at time index $j$ for the $i^{\text{th}}$ unit, the loss function is constructed based on the negative log-likelihood (NLL) of the posterior predictive distribution as:

$$\mathcal{L}_{\text{NLL},ij} = \frac{1}{2}\ln\left(2\pi\left[\sigma_{\text{post},ij}^{(k)}\right]^2\right) + \frac{\left(y_{ij}^{(k)} - \mu_{\text{post},ij}^{(k)}\right)^2}{2\left[\sigma_{\text{post},ij}^{(k)}\right]^2}. \tag{18}$$

Subsequently, the overall loss function across all units and all observation times is obtained as:

$$\mathcal{L}_{\text{NLL}} = \sum_{i=1}^{n}\sum_{j=1}^{m_i}\mathcal{L}_{\text{NLL},ij}. \tag{19}$$

This loss function is adopted because it is directly derived from the Bayesian framework in Eq. (17). By minimizing Eq. (19), the model jointly integrates the population-level prior information from the univariate degradation model and the unit-specific information obtained via the neural networks, which describe the causal influence with uncertainties.

For model training, the Adam optimizer [59] is employed with an initial learning rate $\eta$, and the NLL calculated by Eq. (19) is used as the training objective to learn the network parameters. During training, the NLL is evaluated on the validation set and serves as the validation loss for model selection. A reduce-on-plateau learning rate scheduler is utilized, which halves the learning rate when the validation loss fails to improve for $P$ consecutive epochs, thus preventing premature convergence. Furthermore, an early stopping strategy with a patience of $Q$ epochs is applied to avoid overfitting. The model achieving the lowest validation NLL is retained as the final trained model.

### 2.3 Reliability analysis

In a system characterized by $K$ performance parameters, there exist $K$ corresponding performance thresholds. When the performance of one parameter exceeds its threshold due to degradation, the associated function fails. In this study, we focus on a general system governed by series functional logic, in which the failure of any single function results in overall system failure.

According to belief reliability theory [60], the system reliability is defined as the possibility that the performance margin of the system remains above zero. In this context, the system performance margin $M$ considering competing degradation processes is defined as:

$$M(t) = \min\left(I_{Y_1}\left(Y_1(t) - Y_{1,\text{th}}\right), I_{Y_2}\left(Y_2(t) - Y_{2,\text{th}}\right), \ldots, I_{Y_K}\left(Y_K(t) - Y_{K,\text{th}}\right)\right), \tag{20}$$

where $Y_{i,\text{th}}$ represents the predefined performance threshold of $Y_i$; and $I_{Y_i}$ is an indicator function for $Y_i$, taking a value of -1 when $Y_i$ increases due to degradation, and 1 when $Y_i$ decreases. Notably, in this study, the failure thresholds are treated as deterministic values as simplifying assumptions to focus on modeling the uncertainty in the degradation process itself. If the failure threshold is considered uncertain, the proposed framework can be naturally extended by modeling the failure threshold as a random variable with a specified distribution [61]. In such a case, the performance margin would need to account for two sources of uncertainty simultaneously: the stochastic evolution of the degradation process and the variability in the failure threshold.

Accordingly, the system reliability $R$ can be calculated as:

$$R(t) = \Pr\left\{M(t) > 0\right\}. \tag{21}$$

where Pr denotes the probability measure.

The solution to Eq. (21) is obtained using a Monte Carlo simulation approach. For each simulated unit, the degradation trajectories of the parent performance parameters in the causal graph are first generated. Then, following the causal graph of degradation paths and utilizing the causally dependent degradation modeling method, the degradation of the child node performance parameters is sequentially simulated. This procedure yields the system performance margin for the given unit. By repeating the process multiple runs, a sample set of system performance margins is obtained, which enables the solution to Eq. (21). Without loss of generality, Algorithm 3 outlines the detailed steps of the reliability analysis in the case of two causally dependent degradation processes, where $Y_1$ has a causal influence on $Y_2$.

**Algorithm 3**: Monte Carlo method for reliability analysis of systems with two causally dependent degradation processes.

**Input**:

1. Univariate degradation model parameters obtained via Algorithm 1**:** $\hat{\boldsymbol{\theta}}^{(1)}$ and $\hat{\boldsymbol{\theta}}^{(2)}$.
2. Causal structure obtained via Algorithm 2 (assumed as $Y_1 \rightarrow Y_2$).
3. Trained neural networks $f_\mu(\cdot)$ and $f_\sigma(\cdot)$ based on loss function Eq. (19).
4. Performance thresholds: $Y_{1,\text{th}}$, $Y_{2,\text{th}}$.
5. Simulation time grid $\{t_1,\ldots,t_L\}$.
6. Number of Monte Carlo samples: $B_s$.

**Output**:

1. Estimated system reliability $R(t_l)$ at each time $t_l$.

**Procedure**:

1. **for** $b$ = 1 to $B_s$ **do**
2. Sample initial performance value and degradation rate of $Y_1$ for the $b^{\text{th}}$ simulated unit: $\left(Y_0^{(1)}\right)^{[b]} \sim N\left(\hat{\mu}_{Y_0}^{(1)},\left[\hat{\sigma}_{Y_0}^{(1)}\right]^2\right)$, $\left(a^{(1)}\right)^{[b]} \sim N\left(\hat{\mu}_a^{(1)},\left[\hat{\sigma}_a^{(1)}\right]^2\right)$. Also, sample initial performance value and degradation rate of $Y_2$ for the $b^{\text{th}}$ simulated unit: $\left(Y_0^{(2)}\right)^{[b]} \sim N\left(\hat{\mu}_{Y_0}^{(2)},\left[\hat{\sigma}_{Y_0}^{(2)}\right]^2\right)$.,
$\left(a^{(2)}\right)^{[b]} \sim N\left(\hat{\mu}_a^{(2)},\left[\hat{\sigma}_a^{(2)}\right]^2\right)$.
3. **for** $l$ = 1 to $L$ **do**
4. Generate $y_{bl}^{(1)}$ based on the property of Wiener process via Eq. (2) with $\left(Y_0^{(1)}\right)^{[b]}$ and $\left(a^{(1)}\right)^{[b]}$.
5. Compute prior distribution of $y_{bl}^{(2)}$ based on the univariate degradation model via Eq. (15) with $\left(Y_0^{(2)}\right)^{[b]}$ and $\left(a^{(2)}\right)^{[b]}$.
6. Compute causality distribution of $y_{bl}^{(2)}$ based on the causal influence of $y_{bl}^{(1)}$ using Eq. (14).
7. Compute posterior predictive distribution of $y_{bl}^{(2)}$ using Eq. (17).
8. Sample $y_{bl}^{(2)}$ based on its posterior distribution.
9. Calculate system performance margin $M^{[b]}(t_l)$ via Eq. (20).
10. **end for**
11. **end for**
12. **for** $l$ = 1 to $L$ **do**
13. Calculate reliability $R(t_l)$ based on samples $\left\{M^{[1]}(t_l), M^{[2]}(t_l),\ldots,M^{[B]}(t_l)\right\}$ via Eq. (21).
14. **end for**

2.4 RUL prediction

For a system in operation, one primary concern is determining the time it reaches the failure threshold, i.e., the RUL. At time $t_n$, the degradation data for the $i^{\text{th}}$ performance parameter $\boldsymbol{\xi}_n^{(i)} = \left[\xi_1^{(i)},\xi_2^{(i)},\ldots,\xi_n^{(i)}\right]$ has already been observed. In the univariate degradation model for an individual performance parameter, both the initial performance value $Y_0$ and the degradation rate $a$ are unit-specific. Therefore, by using the observed degradation data $\boldsymbol{\xi}_n^{(i)}$, the parameters associated with $Y_0$ and $a$ can be updated, refining their distributions for more accurate RUL predictions.

For a specific unit, the initial performance value and degradation rate for the $i^{\text{th}}$ performance parameter degradation model are represented as a parameter vector $\boldsymbol{\gamma}^{(i)} = \left[Y_0^{(i)}, a^{(i)}\right]^{\mathrm{T}}$. Given this vector, the observed degradation data follows a Gaussian distribution:

$$\boldsymbol{\xi}_n^{(i)} \left| \boldsymbol{\gamma}^{(i)} \sim N\left(\mathbf{X}^{(i)}\boldsymbol{\gamma}^{(i)}, \boldsymbol{\Sigma}^{(i)}\right)\right. \tag{22}$$

where $\mathbf{X}^{(i)} = [1, \boldsymbol{\Psi}_{\text{obs}}^{(i)}]$, $\boldsymbol{\Psi}_{\text{obs}}^{(i)} = \left[\Psi\left(t_1;\beta^{(i)}\right), \Psi\left(t_2;\beta^{(i)}\right), \ldots, \Psi\left(t_n;\beta^{(i)}\right)\right]$; and $\boldsymbol{\Sigma}^{(i)}$ is a covariance matrix of dimension $n \times n$, with its $(u, v)^{\text{th}}$ element calculated by:

$$\left(\boldsymbol{\Sigma}^{(i)}\right)_{uv} = \left(\sigma^{(i)}\right)^2 \min\left(t_u, t_v\right) + \delta_{uv}\left(\sigma_\varepsilon^{(i)}\right)^2, \tag{23}$$

The prior distribution of $\boldsymbol{\gamma}^{(i)}$ is given by the estimates of the univariate degradation model according to Algorithm 1, which is expressed as:

$$\boldsymbol{\gamma}^{(i)} \sim N\left(\boldsymbol{\mu}_0^{(i)}, \boldsymbol{\Lambda}_0^{(i)}\right) \tag{24}$$

where

$$\boldsymbol{\mu}_0^{(i)} = \left[\hat{\mu}_{Y_0}^{(i)}, \hat{\mu}_a^{(i)}\right]^{\mathrm{T}}, \quad \boldsymbol{\Lambda}_0^{(i)} = \operatorname{diag}\left(\left(\hat{\sigma}_{Y_0}^{(i)}\right)^2, \left(\hat{\sigma}_a^{(i)}\right)^2\right) \tag{25}$$

Under the Bayesian framework, based on Eqs. (22) and (24), the posterior distribution of $\boldsymbol{\gamma}^{(i)}$ considering $\boldsymbol{\xi}_n^{(i)}$ can be derived as:

$$\boldsymbol{\gamma}^{(i)} \left| \boldsymbol{\xi}_n^{(i)} \sim N\left(\boldsymbol{\mu}_{\text{post}}^{(i)}, \boldsymbol{\Lambda}_{\text{post}}^{(i)}\right)\right. \tag{26}$$

where

$$\boldsymbol{\mu}_{\text{post}}^{(i)} = \boldsymbol{\Lambda}_{\text{post}}^{(i)}\left(\left[\boldsymbol{\Lambda}_0^{(i)}\right]^{-1}\boldsymbol{\mu}_0^{(i)} + \left[\mathbf{X}^{(i)}\right]^{\mathrm{T}}\left[\boldsymbol{\Sigma}^{(i)}\right]^{-1}\boldsymbol{\xi}_n^{(i)}\right) \tag{27}$$

$$\boldsymbol{\Lambda}_{\text{post}}^{(i)} = \left(\left[\boldsymbol{\Lambda}_0^{(i)}\right]^{-1} + \left[\mathbf{X}^{(i)}\right]^{\mathrm{T}}\left[\boldsymbol{\Sigma}^{(i)}\right]^{-1}\mathbf{X}^{(i)}\right)^{-1} \tag{28}$$

After updating degradation model parameters, by combining the definition of the system performance margin described by Eq. (20), the RUL of the system can be computed as:

$$T_{\text{RUL}} = \inf\left\{t > t_n \left| M\left(t\right) \le 0\right.\right\} \tag{29}$$

The solution to Eq. (29) is also obtained using a Monte Carlo simulation approach. Notably, when some child node performance parameters are only observable in the laboratory but not in operation, the proposed method allows for RUL prediction by updating solely the distribution of the parent performance parameters. Without loss of generality, Algorithm 4 outlines the detailed steps of the RUL prediction in the case of two causally dependent degradation processes, where $Y_1$ has a causal influence on $Y_2$. The algorithm accounts for scenarios where $Y_2$ is either observable or unobservable during system operation.

**Algorithm 4**: Monte Carlo method for RUL prediction of systems with two causally dependent degradation processes.

**Input**:

1. Univariate degradation model parameters obtained via Algorithm 1**:** $\hat{\boldsymbol{\theta}}^{(1)}$ and $\hat{\boldsymbol{\theta}}^{(2)}$.
2. Observed degradation data up to $t_n$: $\boldsymbol{\xi}_n^{(1)}$ for $Y_1$ and $\boldsymbol{\xi}_n^{(2)}$ for $Y_2$.
3. Causal structure obtained via Algorithm 2 (assumed as $Y_1 \rightarrow Y_2$).
4. Trained neural networks $f_\mu(\cdot)$ and $f_\sigma(\cdot)$ based on loss function Eq. (19).
5. Performance thresholds: $Y_{1,\text{th}}$, $Y_{2,\text{th}}$.
6. Future simulation time grid $\{t_{n+1},\ldots,t_{n+L}\}$.
7. Number of Monte Carlo samples: $B$.

**Output**:

1. Predictive distribution of $T_{\text{RUL}}$.

**Procedure**:

1. **for** $b = 1$ to $B_s$ **do**
2. Obtain posterior distribution of $\boldsymbol{\gamma}^{(1)}$ using $\boldsymbol{\xi}_n^{(1)}$ via Eq. (26).
3. **if** $Y_2$ is observable **do**
4. Obtain posterior distribution of $\boldsymbol{\gamma}^{(2)}$ using $\boldsymbol{\xi}_n^{(2)}$ via Eq. (26).
5. **else do**
6. Set the distribution of $\boldsymbol{\gamma}^{(2)}$ via Eq. (24).
7. **end if**
8. Sample $\left(\boldsymbol{\gamma}^{(1)}\right)^{[b]}$ and $\left(\boldsymbol{\gamma}^{(2)}\right)^{[b]}$ from their distributions.
9. **for** $l = n + 1$ to $n + L$ **do**
10. Generate $y_{t_l}^{(1)}$ based on the property of Wiener process via Eq. (2) with $\left(\boldsymbol{\gamma}^{(1)}\right)^{[b]}$.
11. Compute prior distribution of $y_{t_l}^{(2)}$ based on the univariate degradation model via Eq. (15) with $\left(\boldsymbol{\gamma}^{(2)}\right)^{[b]}$.
12. Compute causality distribution of $y_{t_l}^{(2)}$ based on the causal influence of $y_{t_l}^{(1)}$ using Eq. (14).
13. Compute posterior predictive distribution of $y_{t_l}^{(2)}$ using Eq. (17).
14. Generate $y_{t_l}^{(2)}$ based on its posterior distribution.
16. Compute the system performance margin $M^{[b]}(t_l)$ via Eq. (20).
17. **if** $M^{[b]}(t_l) \leq 0$ **do**
18. $T_{\text{RUL}}^{[b]} \leftarrow t_l - t_n$
19. **break**
20. **end if**
21. **end for**
22. **end for**
23. Obtain the predictive distribution of $T_{\text{RUL}}$ from $\left\{T_{\text{RUL}}^{[1]}, T_{\text{RUL}}^{[2]}, \ldots, T_{\text{RUL}}^{[B]}\right\}$.

## 3 Case study

### 3.1 Dataset description

The effectiveness of the proposed approach is validated using the C-MAPSS dataset [62], provided by NASA's Ames Research Center. It contains sensor measurements collected from several turbofan engines during their operational cycles. As depicted in Fig. 4, a turbofan engine includes a fan, low- and high-pressure compressors, a combustor, high- and low-pressure turbines, and a nozzle. During engine operation, its performance parameters exhibit marked degradation, ultimately leading to engine failure. Since these parameters correspond to different engine performance with inherent physical relationships, the dataset is suitable for validating the proposed methodology.

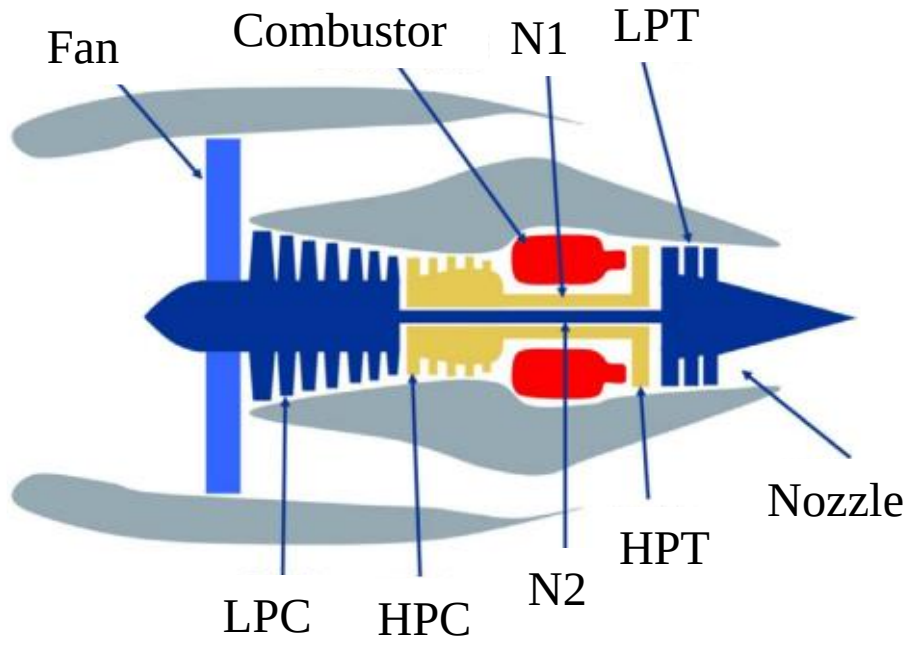

Fig. 4 Diagram of a turbofan engine [62].

Within the C-MAPSS dataset, there are four separate subsets: FD001, FD002, FD003 and FD004, each characterized by different operating conditions and failure patterns. In these datasets, engines in the training sets are run to failure, whereas those in the testing sets are halted before failure occurs. For this case study, the training set of the subset FD001 is utilized, containing run-to-failure sensor data from 100 individual engines. All engines in FD001 share consistent operational conditions and exhibit high-pressure compressor degradation. There are 14 sensor-monitored performance parameters that vary over operational cycles [63]. Among them, this study focuses on two to demonstrate the proposed method: the 8th parameter, "Ratio of fuel flow to the static pressure at high-pressure compressor outlet (pps/psi)," and the 14th, "Low-pressure turbine coolant bleed (lbm/s)," denoted as U8 and U14, respectively. These two parameters are closely related to the degradation of the high-pressure compressor, reflecting fuel supply for combustion and thermal load on the turbine. Fig. 5 illustrates the degradation trajectories of U8 and U14 for the first five engine samples.

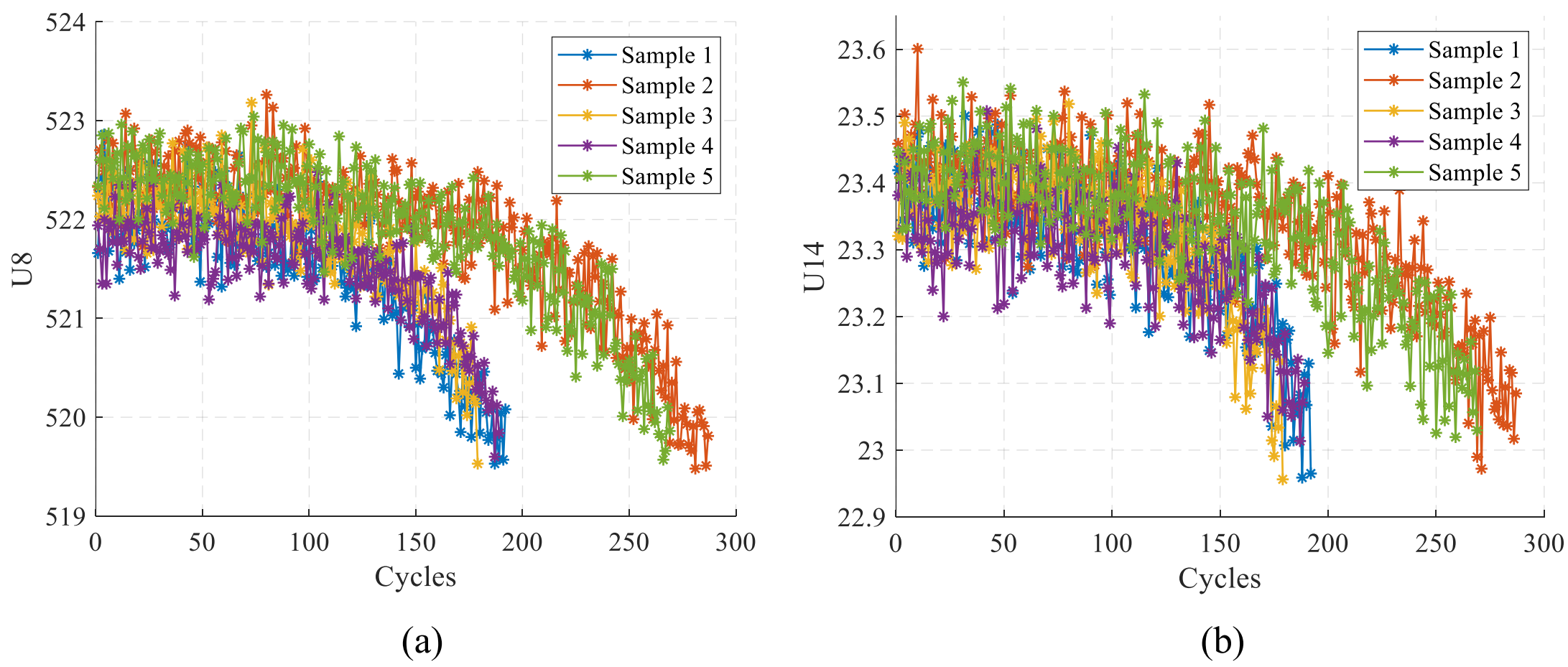

Fig. 5 Degradation trajectories of the first five engine samples: (a) U8; (b) U14.

### 3.2 Univariate degradation modeling

To begin with, the form of time-scale function corresponding to performance degradation should be determined. As analyzed in [62], the time-scale function in Eq. (1) is given by $\Psi(t;\beta)=e^{\beta t}-1$ for the turbofan engine. Then, univariate degradation models of the two performance parameters are developed using Eq. (2). The Wiener process is adopted in this study as a stochastic modeling tool to describe the observed degradation patterns. Notably, the C-MAPSS dataset is generated from a simulation environment, and the observed variables are influenced by controlled modeling assumptions. Therefore, the Wiener process is not intended to represent the true underlying physical degradation mechanism, but rather to capture the temporal variability exhibited in the observed data. This modeling strategy is consistent with existing studies that employ Wiener processes to describe the degradation behaviors in C-MAPSS data [64, 65]. To facilitate subsequent model testing, the first 88 engine samples are used for model development, and the remaining 12 engine samples are used for model testing.

The parameter estimates for the degradation models of the two performance parameters are summarized in Table 3. Fig. 6 illustrates the variation of the log-likelihood values of the two degradation models with respect to the number of iterations of TERIME. It can be observed that the results converge after approximately 500 iterations, which demonstrates the effectiveness of the statistical analysis method using TERIME. Fig. 7 displays the predicted deterministic degradation trends along with the 95% credible intervals derived from the models, along with the observed degradation data. It is evident that the degradation boundaries encompass most of the degradation data, suggesting that the models effectively capture the overall degradation behavior.

To further demonstrate the superiority of the TERIME-based statistical analysis method, comparisons are conducted with the commonly used genetic algorithm (GA) [66] and particle swarm optimization (PSO) [67], and the corresponding results are given in Appendix D.

Table 3 The parameter estimates for the degradation models of the two performance parameters.

| Parameters | U8 | U14 |
|---|---|---|
| $\mu_{Y_0}$ | 521.9175 | 23.3615 |
| $\sigma_{Y_0}$ | 0.4065 | 0.0505 |
| $\mu_a$ | -0.0896 | -0.0119 |
| $\sigma_a$ | 0.0358 | 0.0048 |
| $\beta$ | 0.0178 | 0.0175 |
| $\sigma$ | 0.0125 | 0.0018 |
| $\sigma_\varepsilon$ | 0.3006 | 0.0598 |

Fig. 6 Variation of the log-likelihood values with respect to the number of iterations using TERIME: (a) U8; (b) U14.

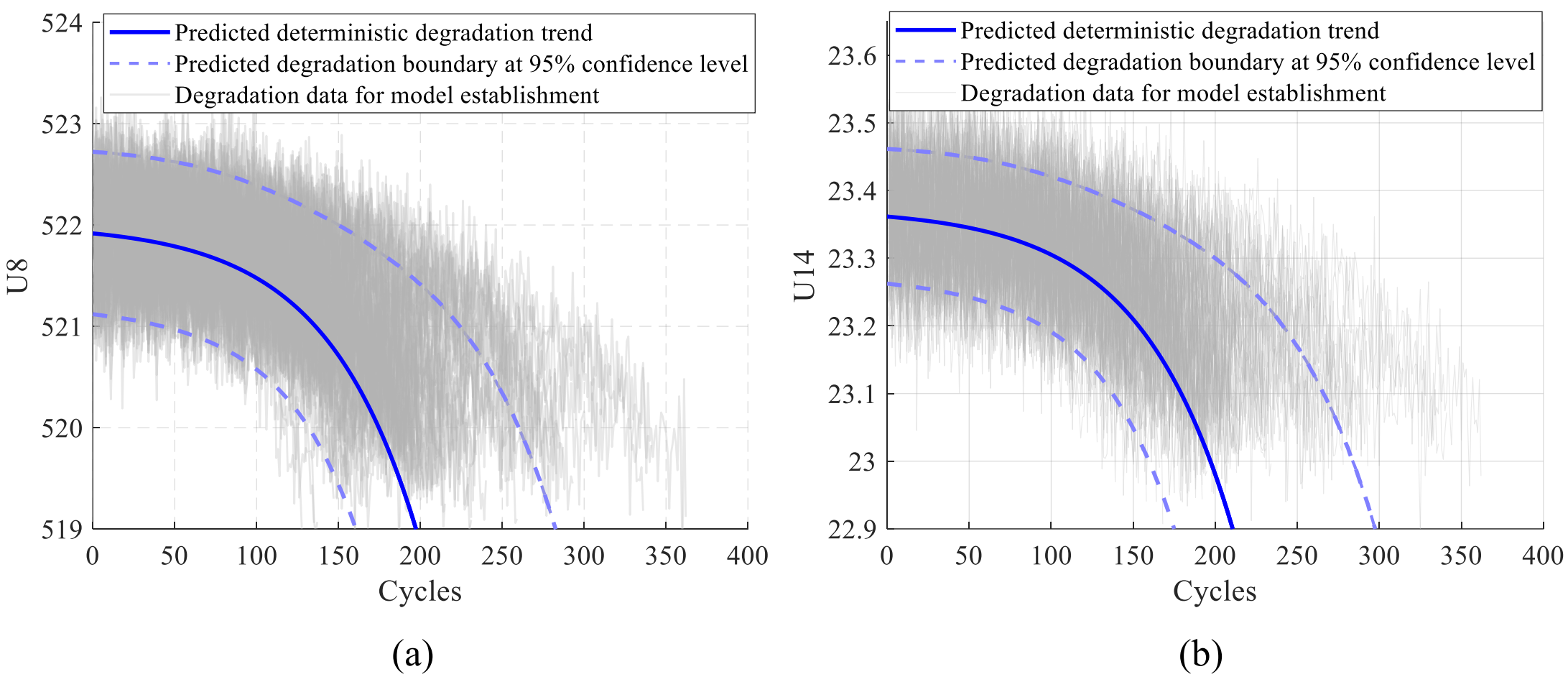

Fig. 7 Predicted deterministic degradation trends and the degradation boundaries at 95% confidence level: (a) U8; (b) U14.

3.3 Causal discovery between degradation processes

According to Algorithm 2, the causal relationship between U8 and U14 can be determined. Notably, measurement noise significantly affects the accuracy of causal discovery. Given the high noise level in the C-MAPSS dataset, the moving average filter is employed to denoise the degradation data, which is a commonly used approach in the literature [68]. Following existing studies [68], the time window for the moving average filter is set to 15. The degradation trajectories of U8 and U14 for the first five engine samples after denoising is shown in Fig. 8.

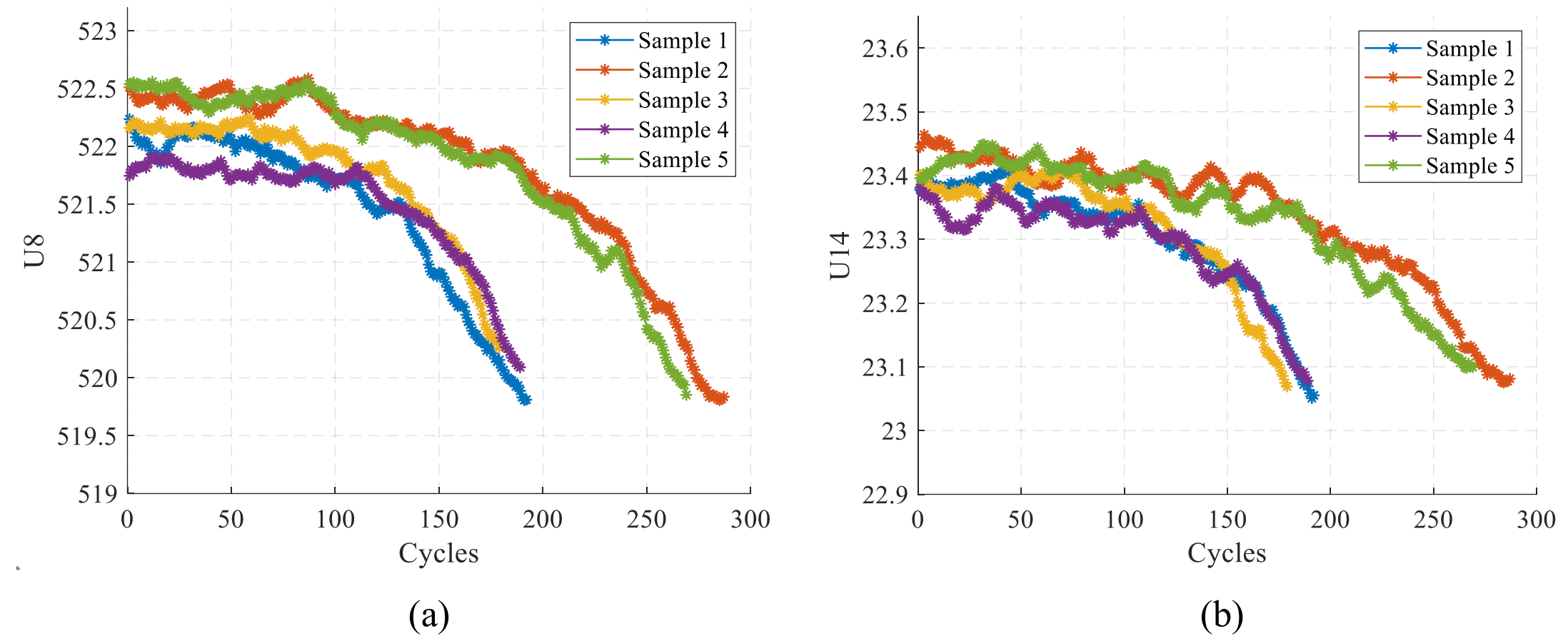

Fig. 8 Degradation trajectories of the first five engine samples after denoising: (a) U8; (b) U14.

To ensure that causal analysis is conducted in the degradation phase, the final 50 cycles of sensor data from each engine are used for causal discovery. Additionally, to enhance result robustness, the causal discovery process is repeated 200 times, each using a randomly selected 80% of the engine samples.

Table 2 displays the number of times a causal relationship is detected in 200 runs of causal discovery. The Stable-PC algorithm consistently identifies a causal link between U8 and U14, with a 100% detection rate, but fails to infer the direction of causality. This indicates the existence of a causal relationship between U8 and U14. However, due to the absence of a V-structure (see Section 2.2.1), the direction of causality cannot be determined. In such cases, domain expertise is required to infer causal direction. In turbofan engines, an increase in U8 affects combustion energy and temperature, which in turn increases the turbine's thermal load, prompting regulation of the coolant bleed represented by U14. Thus, the final causal direction is determined as U8 → U14.

Table 4 Number of times a causal relationship is detected in 200 runs of causal discovery with randomly selected 80% of the engine samples.

| Causal direction | Total times of causal discovery | Times of detected causality | Frequency of detected causality |
|---|---|---|---|
| U8 → U14 | 200 | 200 | 100% |
| U14 → U8 | 200 | 200 | 100% |

Notably, the stable-PC algorithm relies on the significance level $\alpha$ as a key threshold to determine whether a causal connection exists (typically $\alpha$ = 0.05). Fig. 9 shows that in 200 trials, the Fisher $z$- statistic values between U8 and U14 are predominantly less than 1E-5, with the highest values not exceeding 5E-5. This provides evidence against the null hypothesis of independence, indicating a significant causal relationship.

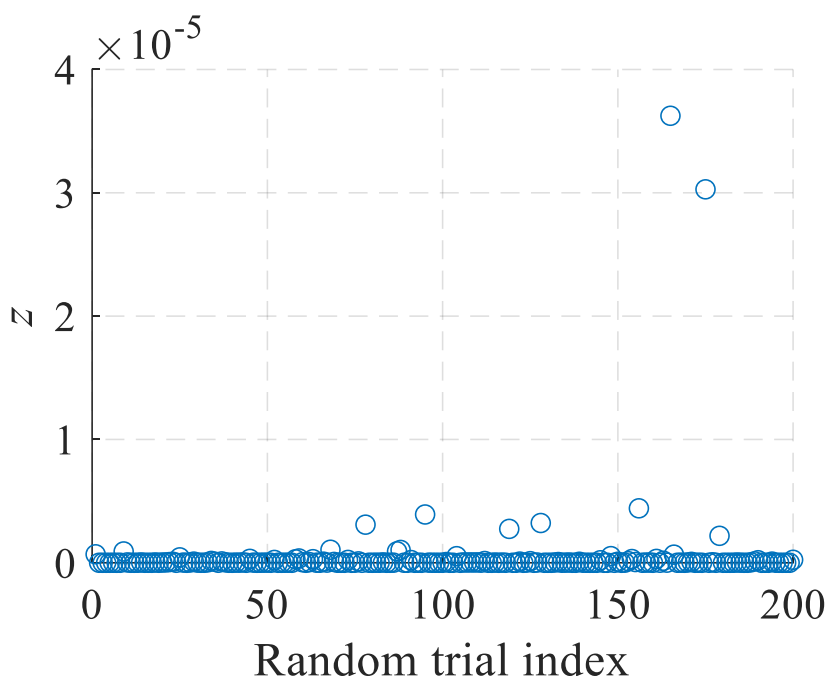

Fig. 9 Fisher z-statistic values between U8 and U14 in different trials.

3.4 Dependent degradation modeling

Based on the parameter estimates of the univariate degradation models in Table 3, an uncertainty-aware neural network is trained by minimizing Eq. (19) to quantify causal effects. The dataset comprises 100 engine degradation samples, of which 88 are used for model development (70 for training and 18 for validation), while the remaining 12 serve as an independent test set. Prior to training, all inputs are normalized using min-max scaling to ensure numerical stability. The network employs a single hidden layer with 4 ReLU-activated neurons and outputs both predictive mean and variance. Training is conducted for up to 1000 epochs using an initial learning rate of 0.001, with a scheduler patience $P$ of 50 and early stopping patience $Q$ of 200.

The training dynamics of the proposed model are illustrated in Fig. 10. The top panel presents the evolution of the training and validation losses. Both curves exhibit a rapid decrease during the initial epochs, followed by a gradual convergence to stable values. The lowest validation loss is achieved at epoch 278, which is marked as the best model for model testing, and the early stopping criterion terminates training before the maximum epoch limit. Vertical dashed lines indicate the learning rate drops triggered by the plateau scheduler, showing that the learning rate is adaptively reduced when no improvement in validation loss is observed. The middle panel reports the gradient norm and parameter norm across epochs. The gradient norm increases sharply in the early stage, then stabilizes around a constant level, indicating steady updates without gradient explosion or vanishing. Meanwhile, the parameter norm grows around the first 60 epochs and subsequently plateaus, suggesting that the model parameters reach a stable scale as training progresses. The bottom panel depicts the learning rate trajectory on a logarithmic scale. Consistent with the scheduler configuration, the learning rate is halved at multiple epochs (e.g., 248, 328 and 378), enabling the optimizer to refine the solution and avoid premature convergence. Overall, the results demonstrate that the model converges stably, with controlled gradient behavior and effective adaptation of the learning rate.

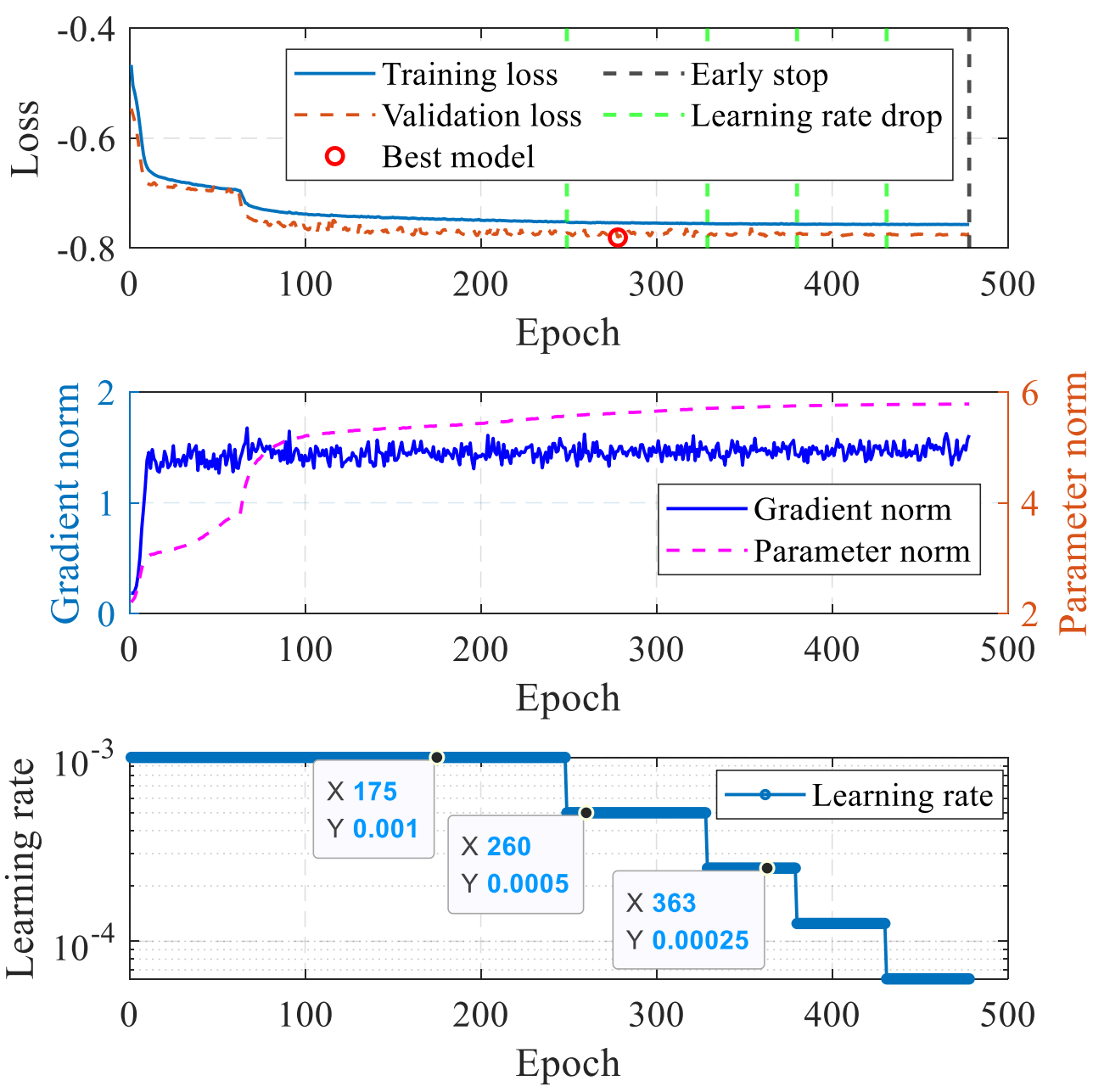

Fig. 10 Training dynamics of the proposed model.

Since the proposed model accounts for the causal dependency between degradation paths, it enables the prediction of the performance degradation of the effect variable using observed degradation data of the cause variable. To this end, U8 from each engine sample in the test set is used as the input to predict the degradation of U14, including both the predicted mean degradation and the 95% credible interval. These predictions are compared with the actual degradation trajectories of U14, as shown in Fig. 11. It can be seen that the predicted trend of U14 mean degradation closely aligns with the actual degradation observations, and the degradation boundary covers the majority of the degradation data, demonstrating the effectiveness of the proposed method.

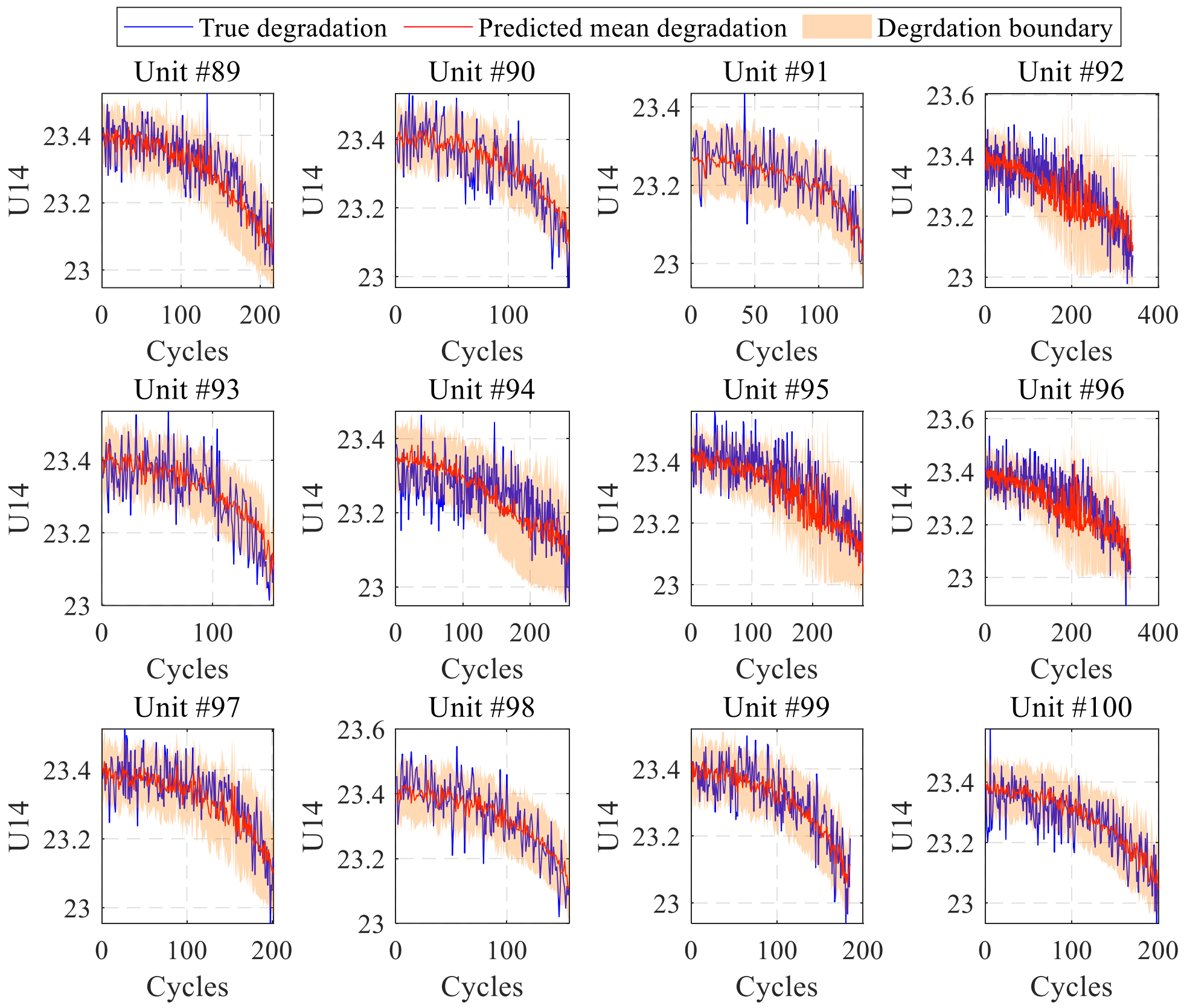

Fig. 11 Predicted mean degradation and the 95% credible interval of U14 on the test set.

### 3.5 Benchmark models and comparisons

#### 3.5.1 Benchmark models

In this subsection, several comparative models are introduced. Firstly, as mentioned in the introduction, correlation-based methods are commonly used for modeling dependent degradation paths. To demonstrate the superiority of the proposed method considering inherent causal relationships, three correlation-based approaches are compared:

- Model $M_1$: A dependent degradation model considering the correlation of initial degradation values and diffusion processes (i.e., Wiener processes) using covariance matrices [16]. Appendix A presents the model form and the parameter estimation results.
- Model $M_2$: A dependent degradation model considering the correlation of initial degradation values and degradation rates (i.e., random effects) using covariance matrices [15]. Appendix B presents the model form and the parameter estimation results.
- Model $M_3$: A dependent degradation model considering the correlation of initial degradation values and degradation increments using the copula method [25]. Based on model selection using Akaike information criterion (AIC), the Frank copula function is chosen. Details regarding the selection of the copula function are provided in the Appendix C.

These three models comprehensively capture the key characteristics of degradation paths, including the correlations between initial performance, degradation rates, degradation increments and degradation fluctuations.

Besides, ablation models are introduced to assess the superiority of combining prior knowledge from the univariate degradation model with observed causal influence in the proposed framework. Two ablation models are considered:

- Model $M_4$: A model assumes independence between U8 and U14, relying only on the univariate degradation model of U14 for prediction, while disregarding causal impact from U8.
- Model $M_5$: A model considers the causal influence of U8 on U14 but ignores the univariate degradation behavior of U14.

### 3.5.2 Evaluation metrics

To quantify the performance of different models in predicting U14 degradation, two types of evaluation metrics are introduced. The first type assesses deterministic prediction accuracy, including mean absolute error (MAE) and root mean square error (RMSE). The formulas for the two metrics are given by:

$$\mathrm{MAE} = \frac{1}{N_s}\sum_{i=1}^{N_s}\left|\hat{y}_i - y_i\right| \tag{30}$$

$$\mathrm{RMSE} = \sqrt{\frac{1}{N_s}\sum_{i=1}^{N_s}\left(\hat{y}_i - y_i\right)^2} \tag{31}$$

where $N_s$ represents the total number of predicted samples; $y_i$ denotes the true degradation value of the $i^{\text{th}}$ predicted sample; and $\hat{y}_i$ indicates the predicted value.

The second type of evaluation metric emphasizes assessing predictive performance in the presence of uncertainty. The negative log-likelihood value $\mathcal{L}_{\mathrm{NLL}}$, as defined in Eq. (19), and the continuous ranked probability score (CRPS) are utilized. The CRPS is defined by:

$$\mathrm{CRPS}\left(F, y_i\right) = \int_{-\infty}^{+\infty}\left[F\left(x\right) - 1\left\{x \geq y_i\right\}\right]^2 \mathrm{d}x \tag{32}$$

where $F(\mathrm{x})$ is the cumulative distribution function of the predictive distribution; and $1\left\{x \geq y_i\right\}$ denotes the indicator function (1 when $x \geq y_i$, and 0 otherwise). For all metrics, lower values indicate better model performance.

### 3.5.3 Comparison results

Given observed degradation data of U8, the degradation of U14 can be predicted based on the conditional probability. A comparison of degradation prediction results for testing units between the correlation-based method and the proposed method is illustrated in Fig. 12, along with the 95% credible interval. Model $M_1$ exhibits significant uncertainty in all degradation predictions and fails to effectively model the degradation process. Model $M_3$ fails to adequately capture the physical relationships between variables, leading to considerable deviations in predicting units with slow degradation rates. Although model $M_2$ achieves the highest accuracy among the correlation-based methods, it demonstrates significant uncertainty when predicting units with slow degradation rates, such as units #92, #94, #95 and #96. In contrast, the proposed method provides accurate degradation predictions for all units, indicating its effectiveness in capturing the dependence between degradation paths.

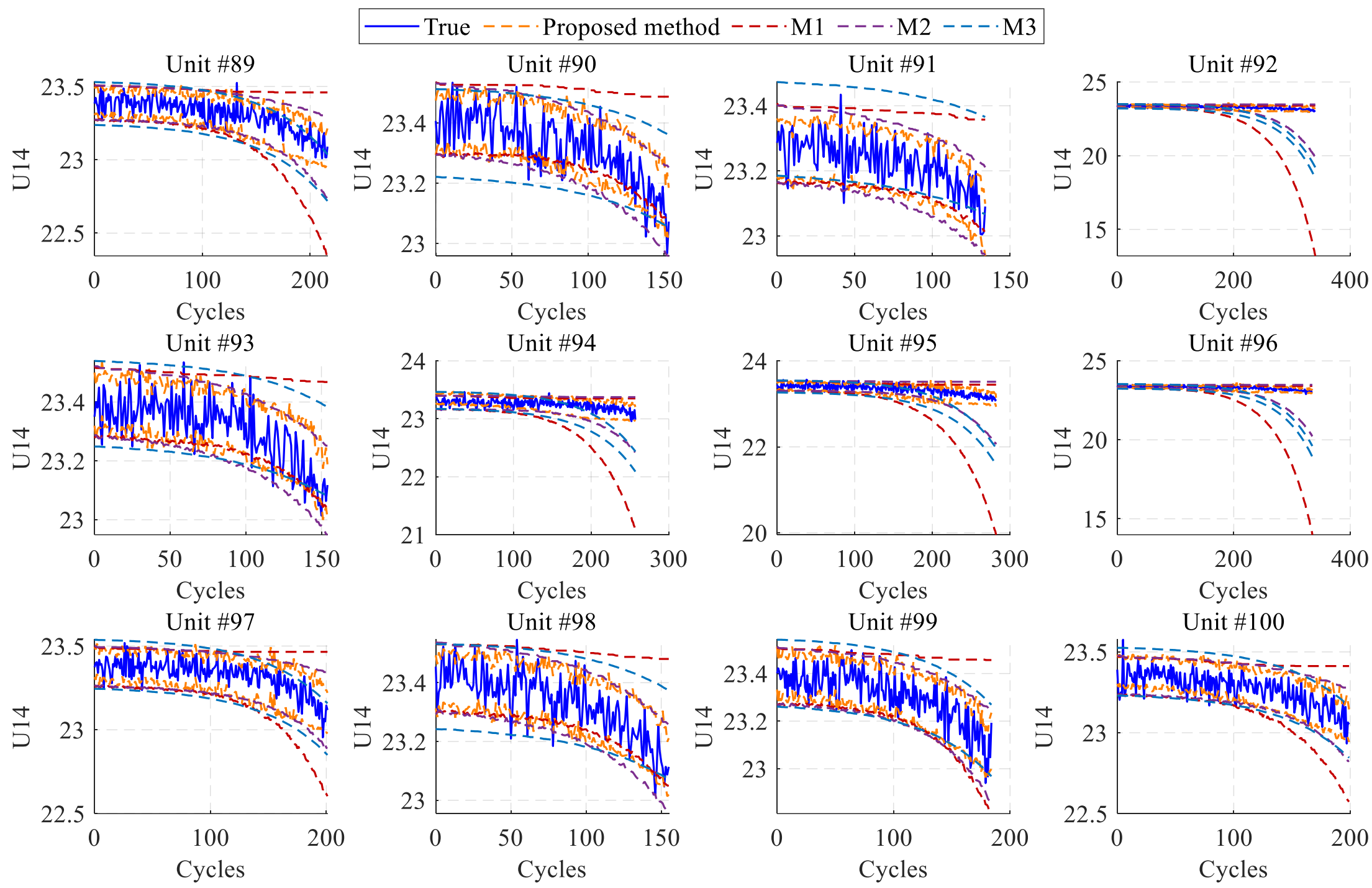

Fig. 12 Comparison of degradation prediction results for testing units between the correlation-based method and the proposed method, along with the 95% credible interval.

Table 5 further presents the overall evaluation results of the correlation-based model for U14 degradation prediction across 12 testing units. The reported RMSE, MAE and CRPS values are computed by pooling all prediction samples from the 12 testing units and evaluating the metrics over the aggregated sample set. It can be seen that the proposed method achieves superior performance in both deterministic degradation prediction and uncertainty quantification, reducing RMSE by 30.2%, MAE by 23.4% and CRPS by 27.6% compared to the best correlation-based approach.

Table 5 Overall evaluation results of the correlation-based model for U14 degradation prediction on 12 testing units.

| Models | RMSE | MAE | CRPS |
|---|---|---|---|
| $M_1$ | 0.607 | 0.248 | 0.165 |
| $M_2$ | 0.106 | 0.077 | 0.058 |
| $M_3$ | 0.612 | 0.250 | 0.162 |
| Proposed model | **0.074** | **0.059** | **0.042** |

***Note***: The best results are highlighted in bold.

Then, Fig. 13 presents the comparison of degradation predictions for testing units from ablation models with actual U14 degradation observations, along with the 95% credible interval. The results show that the model $M_4$ fails to capture the unit-specific degradation behavior, causing large prediction deviations in certain samples, including units #92, #94, #95 and #96. On the other hand, the model $M_5$ lacks prior information about the degradation process, resulting in high uncertainty and broader degradation boundaries for all predictions. Compared to the ablation models, the proposed method achieves superior performance in both deterministic degradation trend prediction and uncertainty quantification.

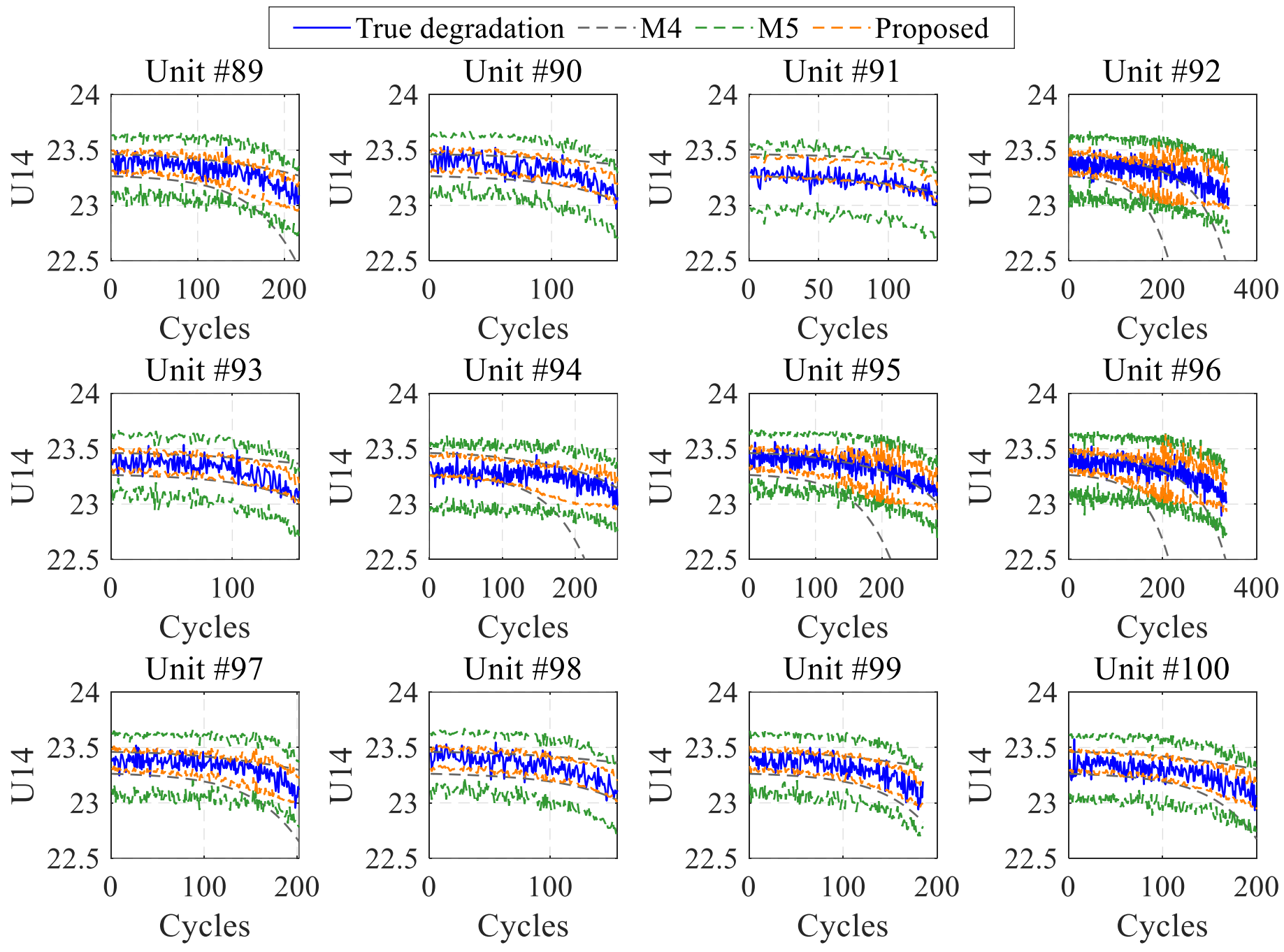

Fig. 13 Comparison of degradation predictions for testing units from ablation and proposed models with actual U14 degradation observations, along with the 95% credible interval.

Table 6 outlines the overall evaluation results of ablation models for U14 degradation prediction on 12 testing units. The reported RMSE, MAE and CRPS values are computed by pooling all prediction samples from the 12 testing units and evaluating the metrics over the aggregated sample set. It is evident that the proposed method surpasses the ablation models in both deterministic degradation trend prediction and uncertainty quantification.

Table 6 Overall evaluation results of ablation models for U14 degradation prediction on 12 testing units.

| Models | RMSE | MAE | $\mathcal{L}_{\text{NLL}}$ | CRPS |
|---|---|---|---|---|
| $M_4$ | 0.619 | 0.259 | -776.41 | 0.191 |
| $M_5$ | 0.076 | 0.060 | -2316.20 | 0.048 |
| Proposed model | **0.074** | **0.059** | **-2867.41** | **0.042** |

***Note***: The best results are highlighted in bold.

### 3.6 Reliability analysis

Based on the developed causally dependent degradation model, the system reliability can be predicted using Algorithm 3. The failure thresholds of U8 and U14 are determined from the observed time-to-failure data, which are 519.6 and 23.07, respectively. Then, the reliability results obtained from the proposed method are compared with those derived from the copula-based dependent degradation model, the independent degradation model considering two variables, and the two univariate degradation models considering single variable, as illustrated in Fig. 14. An earlier decline in system reliability is seen with model $M_1$, where reliability drops under 0.9 at cycle 151. This is caused by the significant uncertainty in its predicted degradation of U14, increasing the likelihood of early failure. Additionally, ignoring the dependency between degradation processes also contributes to an early drop in system reliability, marked by a drop below 0.9 at cycle 152. This results from scenarios where U8 degrades slowly while U14 degrades rapidly, causing premature system failure. By considering the correlations of critical degradation characteristics such as degradation increments and rates, models $M_3$ and $M_2$ show improved reliability predictions, with the reliability falling below 0.9 at cycles 153 and 156, respectively. However, as illustrated in Fig. 12, the degradation of U14 is still occasionally predicted to occur faster than the truth. In contrast, when causal dependencies are integrated, the degradation behavior of U14 is influenced by U8 in individual units, ensuring that slower degradation in U8 also leads to slower degradation in U14. Under such conditions, the proposed model estimates the reliability to drop below 0.9 at cycle 159. This result aligns closely with results obtained when evaluating the failure of only U8 or U14. Fig. 15 displays the failure time differences between U14 and U8 obtained by the proposed method, showing that 50% of the absolute differences are under 3 cycles.

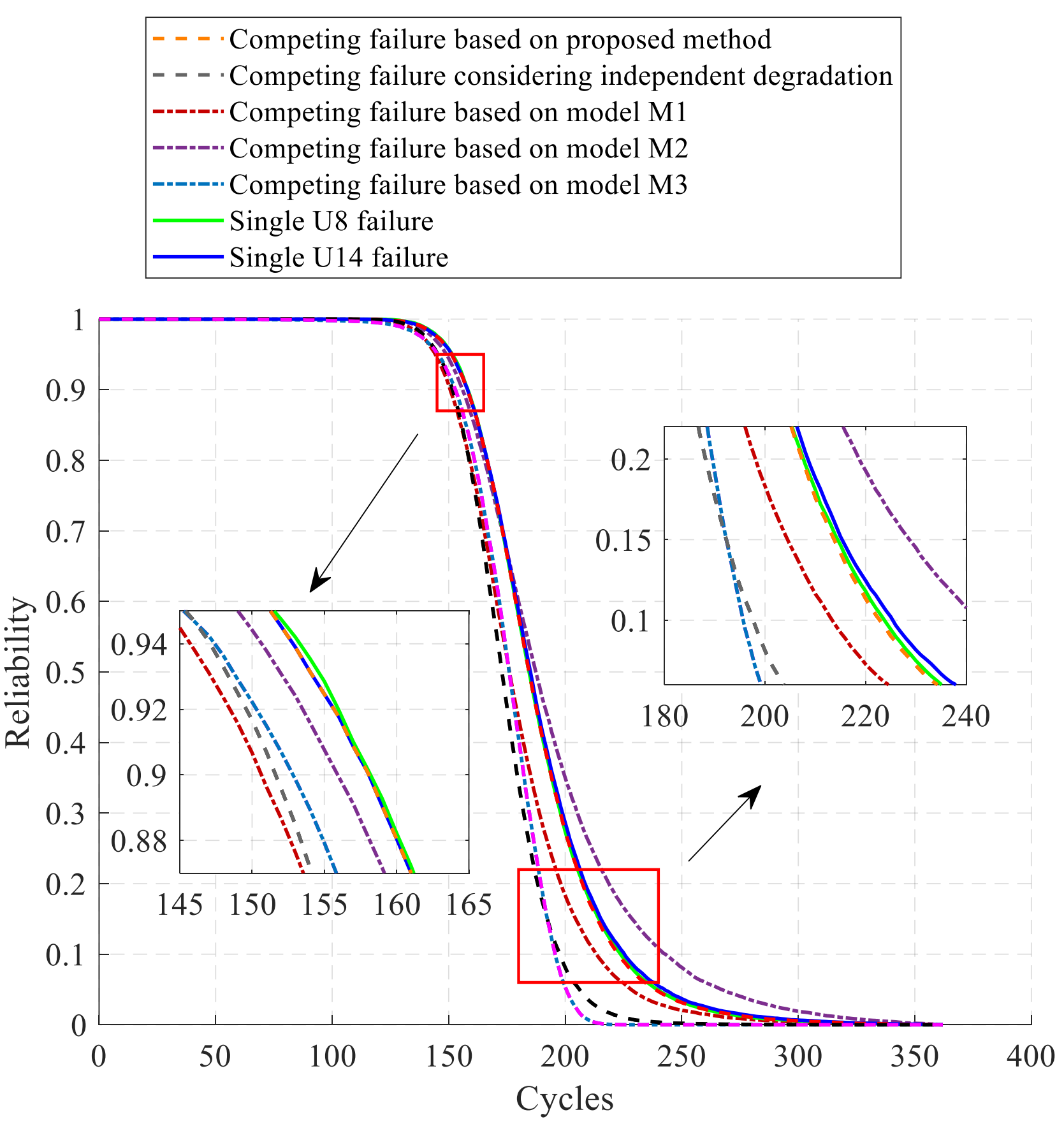

Fig. 14 Comparison of reliability results obtained from different models.

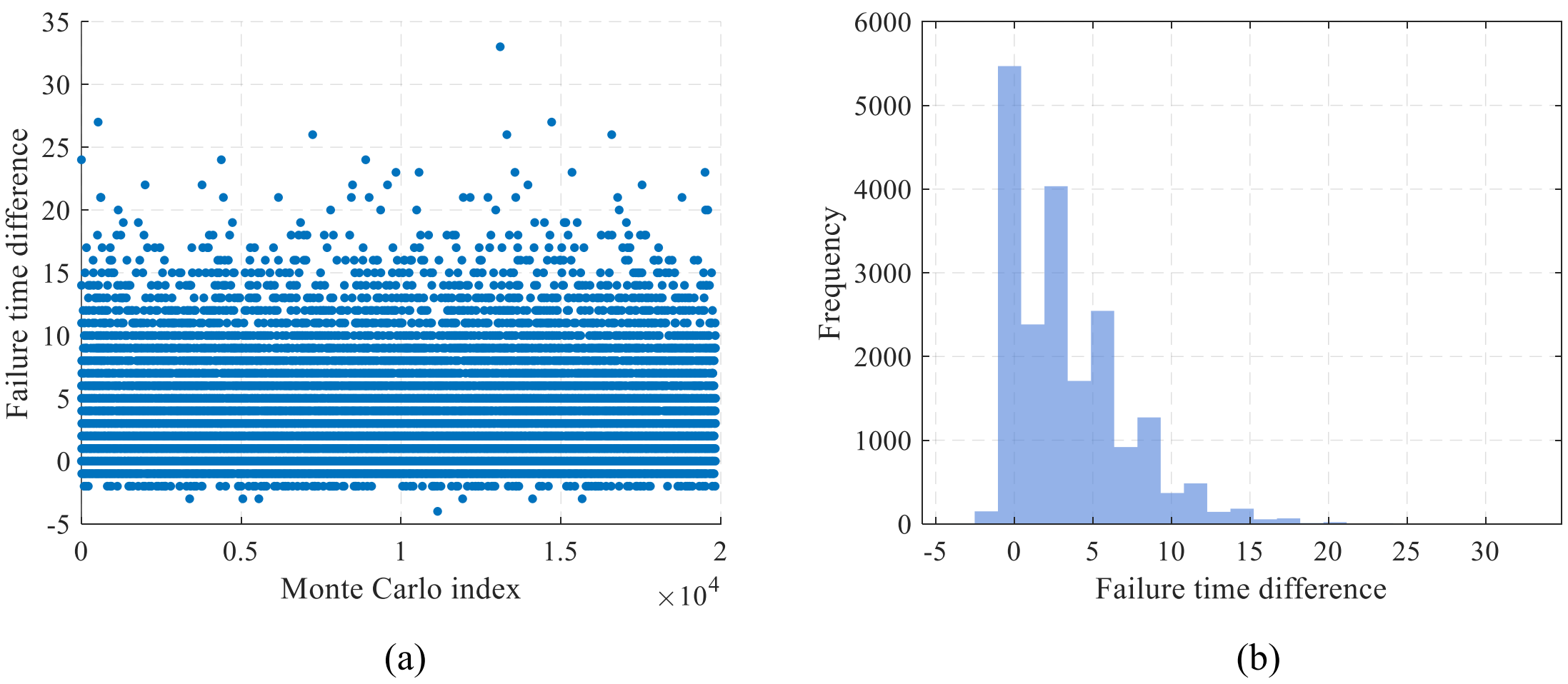

(a) (b)

Fig. 15 Failure time differences between U14 and U8 obtained by the proposed method: (a) individual sample differences; (b) difference distribution.

### 3.7 RUL prediction

#### 3.7.1 Comparison results

According to Algorithm 4, RUL prediction can be carried out for units in operation. The prior steps, which included univariate degradation modeling, causal discovery, causally dependent degradation modeling and reliability analysis, were all performed using the first 88 engine samples. In this section, we focus on predicting the RUL for the remaining 12 samples. Since these test samples fail between 150 and 400 cycles, prediction starts from the 101st cycle. For each test unit, the final denoised values of U8 and U14 are used as failure thresholds. In practice, not all performance indicators can be recorded during system operation. In this study, we assume that U8 is measurable in operation, whereas U14 may be either available or unavailable. When U14 is unmeasurable, RUL prediction considering competing degradation processes can still be performed using the available values of U8.

Table 7 summarizes performance metrics for RUL prediction under U14 measurable and unmeasurable conditions. When U14 is measurable, correlation-based methods fail to surpass the model that assumes independent degradation. This is due to the fact that the observability of both performance indicators allows their past data to include partial dependency, enabling the independent model to perform adequately. Moreover, since correlation does not precisely describe inherent causal relationships, relying on it even slightly impairs prediction accuracy. In contrast, the proposed model, which incorporates causal dependencies, achieves the highest accuracy, with RMSE, MAE and CRPS reduced by approximately 21%, 27%, and 34% compared to the independent model, respectively. When U14 is unmeasurable, the predictive performance of the proposed model remains stable due to its explicit modeling of causal relationships, with variations in all metrics under 0.1%. Conversely, performance of the independent model, $M_1$ and $M_3$ drops considerably, with RMSE, MAE and CRPS increasing by roughly 60% to 80%. Model $M_2$ shows the smallest performance loss, suggesting that the correlation in degradation rates is more significant than that in degradation increments or volatility.

Table 7  Overall evaluation results of RUL predictions on 12 testing units.

| Models | Measurability of U14 | RMSE | MAE | CRPS |
|---|---|---|---|---|
| $M_1$ | Yes | 52.40 | 36.27 | 30.37 |
| $M_2$ | Yes | 54.07 | 39.38 | 33.11 |
| $M_3$ | Yes | 51.36 | 35.34 | 29.43 |
| $M_4$ | Yes | 50.88 | 34.48 | 29.03 |
| Proposed model | Yes | **40.16** | **25.19** | **19.04** |
| $M_1$ | No | 83.95 | 66.26 | 51.51 |
| $M_2$ | No | 66.25 | 53.36 | 41.31 |
| $M_3$ | No | 81.30 | 62.39 | 55.37 |
| $M_4$ | No | 80.42 | 61.86 | 51.04 |
| Proposed model | No | **40.17** | **25.21** | **19.05** |

***Note***: The best results under each scenario are highlighted in bold.

Fig. 16 compares the predicted and actual RUL values for all test units using different methods. It can be seen that when U14 is unmeasurable, the independent and correlation-based models fail to provide accurate RUL predictions for test samples with slower degradation, such as units #92, #94, #95 and #96. This aligns with the large degradation prediction deviations observed in Figs 12 and 13 for these models. By comparison, the proposed model stands out by offering precise RUL predictions in both measurable and unmeasurable conditions.

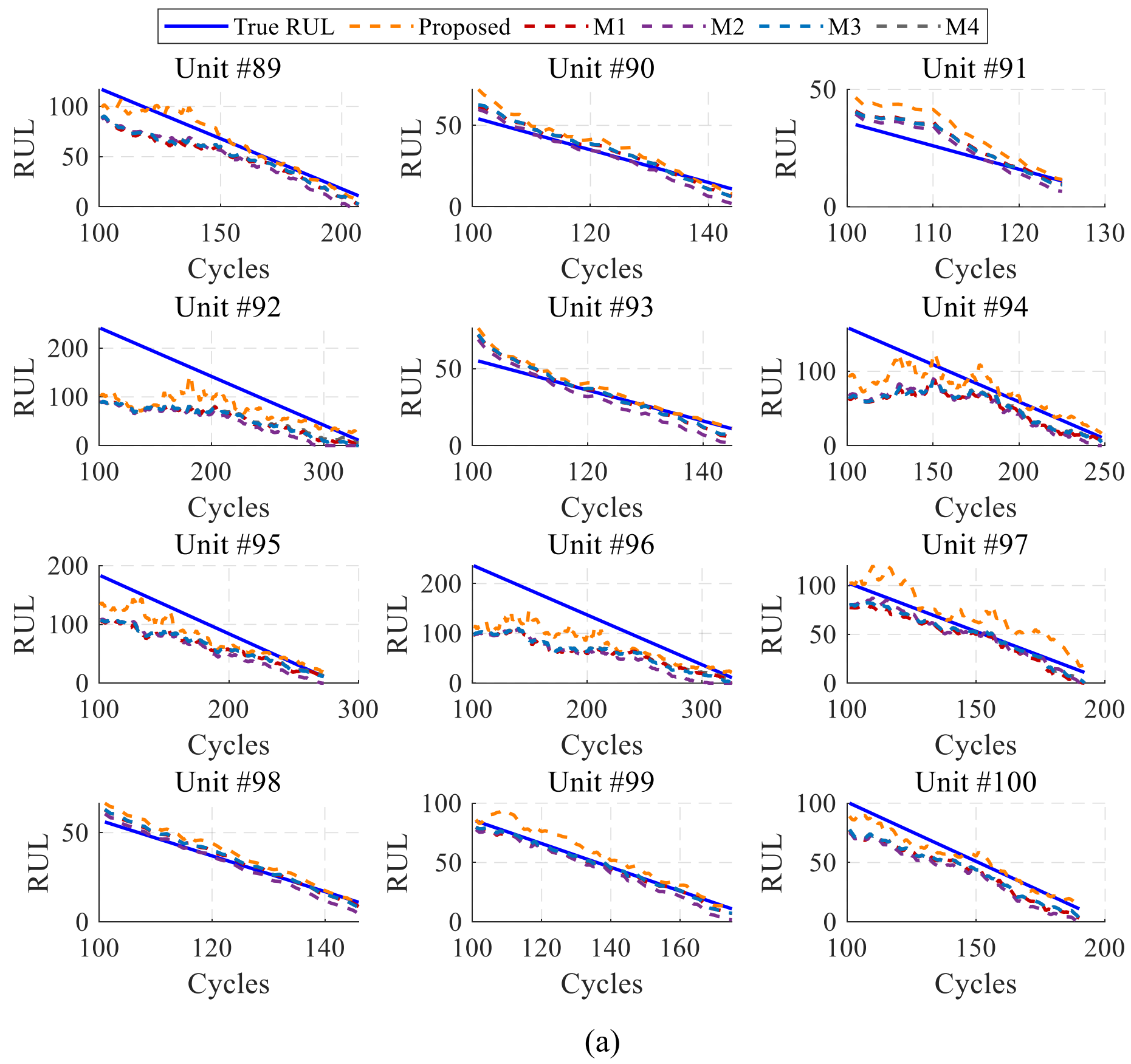

(a)

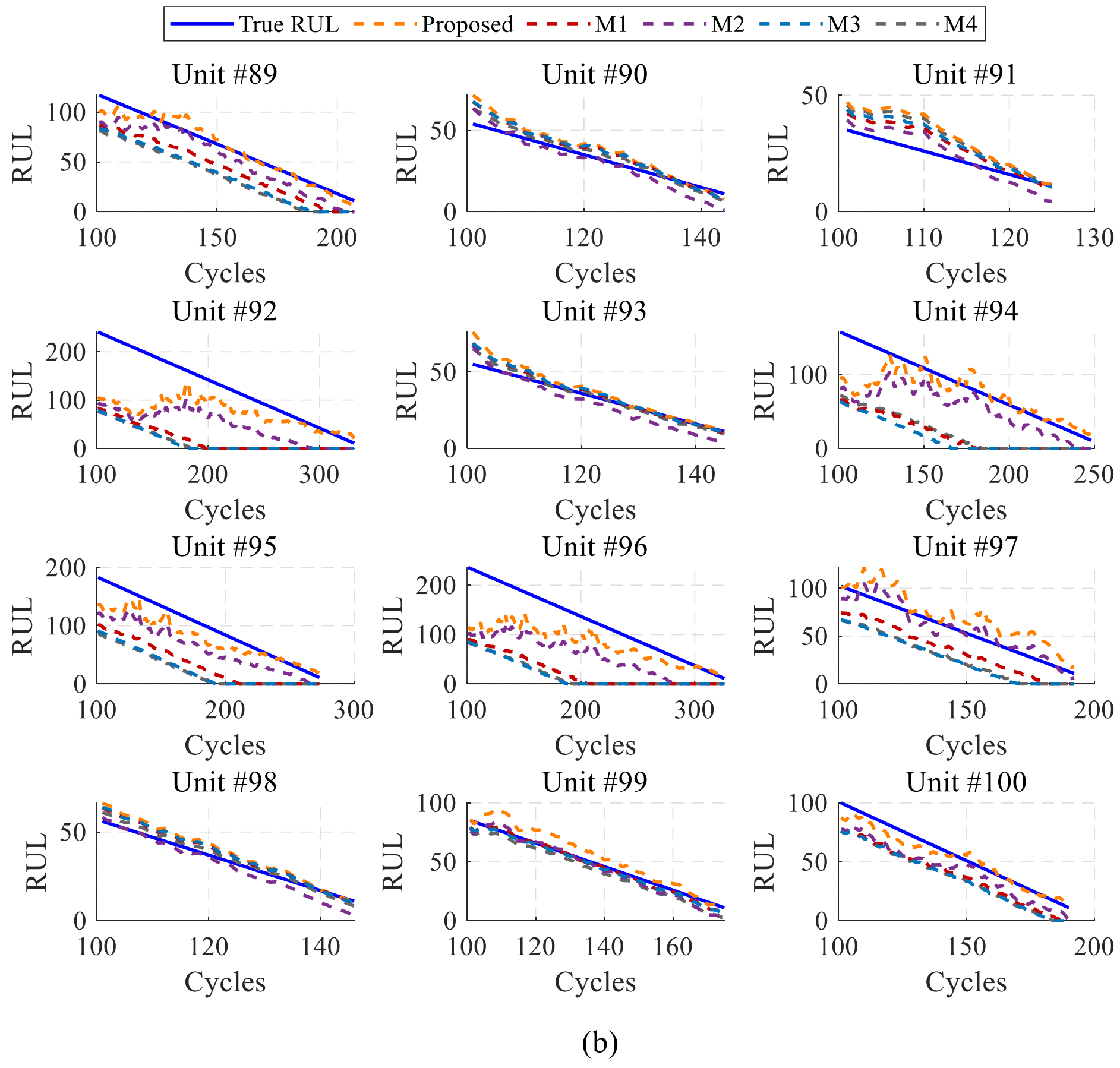

(b)

Fig. 16 Comparison between predicted and true RUL values for test units using different methods: (a) U14 is observable; (b) U14 is unobservable.

#### 3.7.2 Discussions

Firstly, to investigate the impact of hyper-parameters in the uncertainty-aware neural network on model performance, additional experiments are conducted by varying the learning rate and the number of hidden neurons, as summarized in Table 8. The results show that the model performance remains relatively stable across different configurations, with only marginal variations in RMSE, MAE and CRPS. Specifically, changing the learning rate from 0.001 to 0.01 leads to negligible differences in prediction accuracy, and increasing the number of hidden neurons from 4 to 16 does not yield significant performance improvements. This indicates that the proposed model is robust to hyper-parameter settings.

Table 8 Effect of learning rate and hidden neuron number on RUL prediction performance.

| Learning rate | Number of hidden neurons | RMSE | MAE | CRPS |
|---|---|---|---|---|
| 0.001 | 4 | 40.16 | 25.19 | 19.04 |
| 0.005 | 4 | 40.43 | 25.34 | 19.19 |
| 0.01 | 4 | 40.04 | 25.17 | 19.03 |
| 0.001 | 8 | 40.15 | 25.20 | 19.04 |
| 0.001 | 16 | 40.28 | 25.26 | 19.12 |

Furthermore, the causal direction is determined by domain knowledge in this study. To examine the impact of causal direction on model performance, an additional experiment is conducted by reversing the identified relationship between U8 and U14 (i.e., assuming U14 → U8). The corresponding results are compared with those obtained under the original direction U8 → U14. The results are listed in Table 9. It is observed that the reversed direction leads to degraded predictive performance, indicating that the original direction U8 → U14 is more consistent with the underlying physics. Nevertheless, it is worth noting that even under the reversed direction, the proposed framework still outperforms traditional correlation-based models given in Table 7. This suggests that incorporating causal structure, even when the causal direction is partially uncertain, provides a more informative modeling framework than purely correlation-based approaches. Therefore, although domain knowledge may cause ambiguity, the proposed method remains robust and demonstrates clear advantages over existing correlation-based degradation models.

Table 9 Effect of causal direction on RUL prediction performance.

| Causal direction | RMSE | MAE | CRPS |
|---|---|---|---|
| U8 → U14 | 40.16 | 25.19 | 19.04 |
| U14 → U8 | 41.68 | 28.23 | 21.11 |

## 4 Conclusion

This paper focuses on modeling dependent degradation data for reliability analysis and remaining useful life prediction in systems characterized by multiple performance indicators. To address the limitations of traditional correlation-based modeling approaches, a causality-driven degradation modeling framework is proposed, integrating stochastic process-based univariate degradation models and causal dependencies between degradation paths within a unified framework. Based on the case study using the C-MAPSS dataset, the following conclusions can be drawn:

- Given the degradation trajectories of the parent performance parameter U8, the proposed model effectively predicts the degradation trajectories of the child performance parameter U14, achieving reductions of approximately 30%, 23% and 27% in RMSE, MAE and CRPS compared to the best correlation-based model. In addition, excluding univariate degradation while solely considering causal influence leads to increased uncertainty in degradation predictions, with a 14% rise in CRPS, highlighting the effectiveness of integrating causal dependencies with the stochastic process-based univariate degradation model.
- The proposed method yields different reliability analysis results by eliminating physically unrealistic scenarios where U8 degrades slowly while U14 degrades rapidly and triggers system failure.
- The proposed method, which integrates inherent causal relationships, outperforms other models in predicting RUL, achieving reductions of roughly 22%, 29% and 35% compared to the best correlation-based model. Furthermore, when U14 is unmeasurable, the proposed model achieves almost consistent RUL prediction results, while other methods undergo a notable reduction in predictive effectiveness.

These findings indicate that, compared to correlation-based methods, the proposed method that incorporates causal relationships provides a more accurate description of the dependencies between degradation paths. Beyond the work of this study, several directions warrant further investigation. Firstly, the proposed framework is developed under Gaussian assumptions. Its extension to non-Gaussian settings is a meaningful direction for future work. Specifically, the Wiener process can be generalized to alternative stochastic processes, such as Gamma or inverse Gaussian processes, to better capture non-Gaussian degradation behaviors. The conditional independence tests used in Stable-PC can be replaced by nonparametric methods that do not rely on distributional assumptions. The degradation information fusion strategy can be extended using approximate inference techniques to obtain the posterior distribution, such as Monte Carlo sampling or variational inference. These extensions would further enhance the flexibility and applicability of the proposed framework. Second, the proposed framework is primarily developed for degradation data with relatively dense observations, where sufficient samples are available for degradation modeling and causal discovery. In scenarios with sparse data, the performance of the method may be affected due to limited information. In such cases, transfer learning may help improve model performance under limited data conditions.

## Declarations of interest

None.

## Acknowledgements

This work was supported by the Natural Science Foundation of Beijing Municipality [grant number L252210].

## Appendix A

Based on the univariate degradation model in Eq. (2), this appendix presents the formulation of the dependent degradation model that incorporates correlation between initial performance values and Wiener processes, denoted as $M_1$.

For $k$ = 1, 2, the degradation process is defined as:

$$\begin{aligned} &Y_k(t) = Y_0^{(k)} + a^{(k)}\Psi\left(t;\beta^{(k)}\right) + \sigma^{(k)}B_k(t) + \varepsilon^{(k)}, \\ &a^{(k)} \sim N\left(\mu_a^{(k)}, \left[\sigma_a^{(k)}\right]^2\right), \varepsilon^{(k)} \sim N\left(0, \left[\sigma_\varepsilon^{(k)}\right]^2\right), \end{aligned} \tag{33}$$

where $i^{(n)}$ represents the parameter $i$ in the univariate degradation model of the $n^{\text{th}}$ performance parameter.

The dependence between the initial performance values is modeled through a bivariate normal distribution:

$$\begin{pmatrix} Y_0^{(1)} \\ Y_0^{(2)} \end{pmatrix} \sim N\left(\begin{pmatrix} \mu_{Y_0}^{(1)} \\ \mu_{Y_0}^{(2)} \end{pmatrix}, \mathbf{\Sigma}_0\right), \tag{34}$$

with covariance matrix

$$\mathbf{\Sigma}_0 = \begin{pmatrix} \left[\sigma_{Y_0}^{(1)}\right]^2 & \rho_0 \sigma_{Y_0}^{(1)} \sigma_{Y_0}^{(2)} \\ \rho_0 \sigma_{Y_0}^{(1)} \sigma_{Y_0}^{(2)} & \left[\sigma_{Y_0}^{(2)}\right]^2 \end{pmatrix}. \tag{35}$$

Here, $\rho_0 \in (-1,1)$ denotes the correlation coefficient between the initial performance values.

Temporal dependence between the two degradation paths is further considered through correlated Wiener processes as:

$$\begin{pmatrix} B_1(t) \\ B_2(t) \end{pmatrix} \sim N(0, \mathbf{\Sigma}_B), \tag{36}$$

with covariance matrix

$$\mathbf{\Sigma}_B = \begin{pmatrix} \left[\sigma^{(1)}\right]^2 & \rho_B \sigma^{(1)} \sigma^{(2)} \\ \rho_B \sigma^{(1)} \sigma^{(2)} & \left[\sigma^{(2)}\right]^2 \end{pmatrix}. \tag{37}$$

Here, $\rho_B \in (-1,1)$ denotes the correlation coefficient between the Wiener processes. Model $M_1$ is fully defined by Eqs. (33)-(37).

The model parameters are estimated using TERIME. The parameter estimates are summarized in Table 10.

Table 10 The parameter estimates for the model $M_1$.

| Parameters | Values | Parameters | Values |
|---|---|---|---|
| $\mu_{Y_0}^{(1)}$ | 521.9162 | $\mu_{Y_0}^{(2)}$ | 23.3603 |
| $\sigma_{Y_0}^{(1)}$ | 0.3713 | $\sigma_{Y_0}^{(2)}$ | 0.0462 |
| $\mu_a^{(1)}$ | -0.0665 | $\mu_a^{(2)}$ | -0.0094 |
| $\sigma_a^{(1)}$ | 0.0462 | $\sigma_a^{(2)}$ | 0.0059 |
| $\beta^{(1)}$ | 0.0183 | $\beta^{(2)}$ | 0.0181 |
| $\sigma^{(1)}$ | 0.0120 | $\sigma^{(2)}$ | 0.0016 |
| $\sigma_\varepsilon^{(1)}$ | 0.3006 | $\sigma_\varepsilon^{(2)}$ | 0.0598 |
| $\rho_0$ | 0.9941 | $\rho_B$ | 0.9934 |

## Appendix B

This appendix presents the formulation of the dependent degradation model that incorporates correlation between initial performance values and degradation rates, denoted as $M_2$.

For $k$ = 1, 2, the degradation process is defined as:

$$Y_k(t) = Y_0^{(k)} + a^{(k)} \Psi\left(t; \beta^{(k)}\right) + \sigma^{(k)} B_k(t) + \varepsilon^{(k)}, \varepsilon^{(k)} \sim N\left(0, \left[\sigma_\varepsilon^{(k)}\right]^2\right), \tag{38}$$

The dependence between the initial performance values is the same as that in model $M_1$, described by Eqs. (34) and (35). The dependence between the degradation rates is modeled through a bivariate

normal distribution:

$$\begin{pmatrix} a^{(1)} \\ a^{(2)} \end{pmatrix} \sim N\left( \begin{pmatrix} \mu_a^{(1)} \\ \mu_a^{(2)} \end{pmatrix}, \mathbf{\Sigma}_a \right), \tag{39}$$

with covariance matrix

$$\mathbf{\Sigma}_a = \begin{pmatrix} \left[\sigma_a^{(1)}\right]^2 & \rho_a \sigma_a^{(1)} \sigma_a^{(2)} \\ \rho_a \sigma_a^{(1)} \sigma_a^{(2)} & \left[\sigma_a^{(2)}\right]^2 \end{pmatrix}. \tag{40}$$

Here, $\rho_a \in (-1,1)$ denotes the correlation coefficient between the degradation rates. Model $M_2$ is fully defined by Eqs. (34), (35) and (38)-(40).

The model parameters are estimated using TERIME. The parameter estimates are summarized in Table 11.

Table 11 The parameter estimates for the model $M_2$.

| Parameters | Values | Parameters | Values |
|---|---|---|---|
| $\mu_{Y_0}^{(1)}$ | 521.9157 | $\mu_{Y_0}^{(2)}$ | 23.3609 |
| $\sigma_{Y_0}^{(1)}$ | 0.3794 | $\sigma_{Y_0}^{(2)}$ | 0.0474 |
| $\mu_a^{(1)}$ | -0.0664 | $\mu_a^{(2)}$ | -0.0092 |
| $\sigma_a^{(1)}$ | 0.0424 | $\sigma_a^{(2)}$ | 0.0055 |
| $\beta^{(1)}$ | 0.0184 | $\beta^{(2)}$ | 0.0183 |
| $\sigma^{(1)}$ | 0.0115 | $\sigma^{(2)}$ | 0.0013 |
| $\sigma_\varepsilon^{(1)}$ | 0.3007 | $\sigma_\varepsilon^{(2)}$ | 0.0598 |
| $\rho_0$ | 0.9986 | $\rho_B$ | 0.9993 |

## Appendix C

In this appendix, we introduce copula functions and the method used to select the most suitable one. According to [25], the bivariate copulas considered include Gaussian copula, Student's *t* copula, Frank copula, Clayton copula and Gumbel copula. Their corresponding copula density functions are listed in Table 12. To compare alternative copula methods, we employ the AIC defined by:

$$\mathrm{AIC} = 2p - 2\hat{L}, \tag{41}$$

where $p$ is the total number of parameters and $\hat{L}$ is the value of the likelihood function for the fitted model. Given two candidate competing copulas, the model with the lower AIC value is preferred.

Table 12 Candidate copula methods and corresponding copula density functions.

| Methods | Density function |
|---|---|
| Gaussian | $c(u,v;\theta)=\frac{1}{\sqrt{1-\theta^2}}\exp\left(-\frac{\theta^2\left(\left[\Phi^{-1}(u)\right]^2+\left[\Phi^{-1}(v)\right]^2\right)-2\theta\,\Phi^{-1}(u)\,\Phi^{-1}(v)}{2\left(1-\theta^2\right)}\right)$ |
| Student's $t$ | $c(u,v;\theta,\rho)=\frac{\Gamma\left(\frac{\theta+2}{2}\right)\Gamma\left(\frac{\theta}{2}\right)}{\Gamma\left(\frac{\theta+1}{2}\right)^2\sqrt{1-\rho^2}}\left[1+\frac{x^2-2\rho xy+y^2}{\theta(1-\rho^2)}\right]^{-\frac{\theta+2}{2}}\left(1+\frac{x^2}{\theta}\right)^{\frac{\theta+1}{2}}\left(1+\frac{y^2}{\theta}\right)^{\frac{\theta+1}{2}}$, $x=t_\theta^{-1}(u), y=t_\theta^{-1}(v)$ |
| Frank | $c(u,v;\theta)=\frac{\theta\left(1-e^{-\theta}\right)e^{-\theta(u+v)}}{\left[1-e^{-\theta}-\left(1-e^{-\theta u}\right)\left(1-e^{-\theta v}\right)\right]^2}$ |
| Clayton | $c(u,v;\theta)=(\theta+1)(uv)^{-(\theta+1)}\left(u^{-\theta}+v^{-\theta}-1\right)^{-(2+1/\theta)}$ |
| Gumbel | $c(u,v;\theta)=\exp\left(-A^{1/\theta}\right)\frac{(\ln u\ln v)^{\theta-1}}{uv}A^{1/\theta-2}\left(\theta-1+A^{1/\theta}\right), A=\left[(-\ln u)^\theta+(-\ln v)^\theta\right]$ |

For the case study, the dependence of degradation increments and initial values for U8 and U14 are evaluated using the candidate copula methods, and the corresponding results are summarized in Table 13. Among all candidate models, the Frank copula consistently yields the lowest AIC values for both types of dependence. Therefore, the Frank copula is selected as the most suitable model for the analysis.

Table 13 AIC comparison of copula methods for the case study.

| Methods | Dependence of degradation increment | Dependence of initial performance |
|---|---|---|
| Gaussian | -3.764 | -16.472 |
| Student's $t$ | -1.853 | -14.473 |
| Clayton | 0.201 | -9.992 |
| Frank | -4.238 | -16.958 |
| Gumbel | -1.371 | -14.797 |

## Appendix D

In this appendix, the performance of TERIME is compared with GA and PSO for solving the parameter estimation problem of the univariate degradation model in Eq. (7). Fig. 17 shows the evolution of the log-likelihood values for the two degradation models as the number of iterations increases under different optimization methods. The results indicate that TERIME attains significantly higher log-likelihood values than GA and PSO, suggesting improved optimization performance. This improvement can be attributed to a more effective balance between global exploration and local exploitation, which facilitates more accurate identification of the optimal solution.

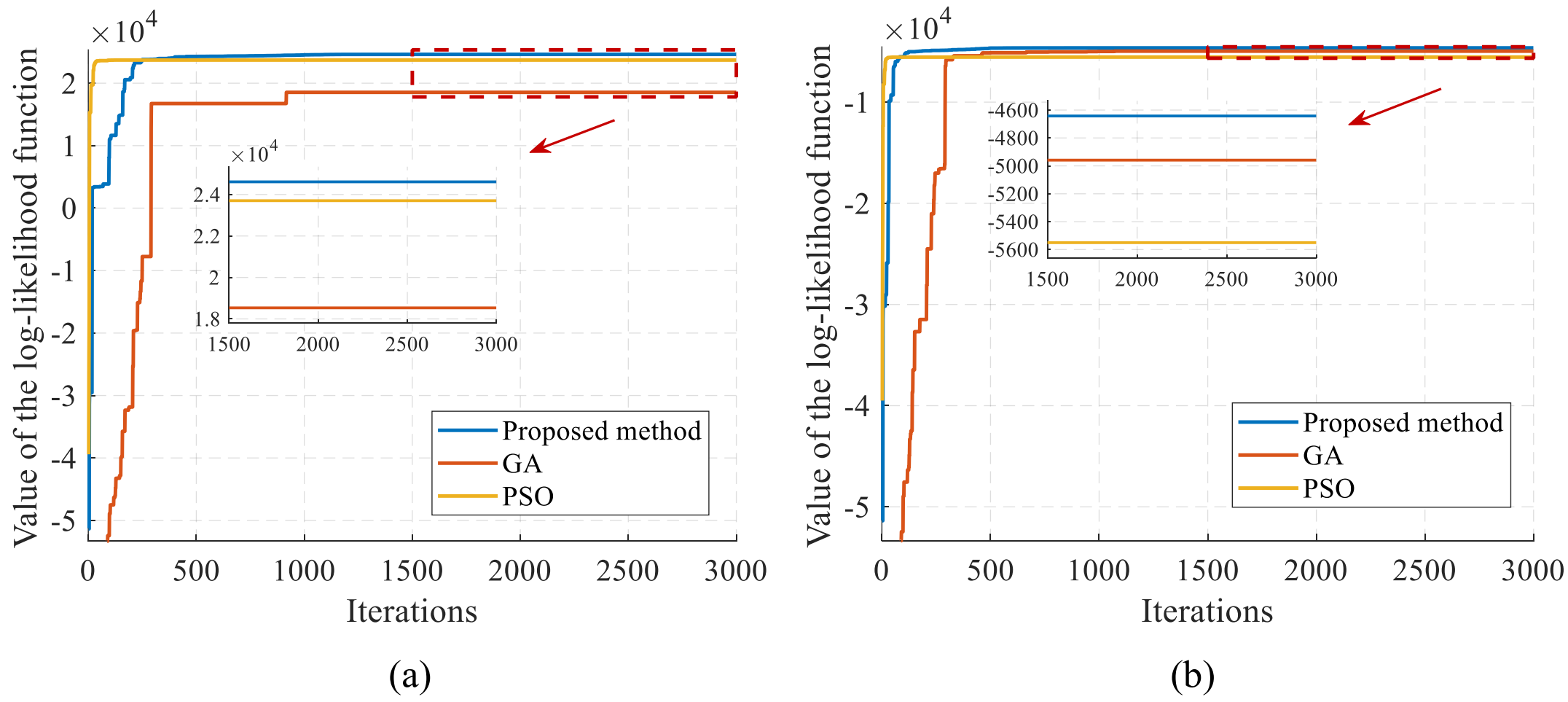

Fig. 17 Variation of the log-likelihood values with respect to the number of iterations using different algorithms: (a) U8; (b) U14.